\documentclass[
reprint,
superscriptaddress,
nofootinbib,
 amsmath,amssymb,
 aps,
 prl,
twocolumn,
]{revtex4-2}

\usepackage[pdftex]{graphicx}
\usepackage{epsfig} 
\usepackage{dcolumn}
\usepackage{bm}
\usepackage{hyperref}
\usepackage{color}
\usepackage[normalem]{ulem}
\usepackage{upgreek}
\usepackage{overpic}
\usepackage[mathlines]{lineno}
\usepackage[utf8]{inputenc}
\usepackage{orcidlink}
\newcounter{Req}
\newcounter{Raux}

  \newenvironment{Reqnarray}
  {\setcounter{Raux}{\theequation}
    \setcounter{equation}{\theReq}%
    \renewcommand\theequation{R\arabic{equation}}
    \eqnarray}
  {\endeqnarray\setcounter{Req}{\value{equation}}\setcounter{equation}{\theRaux}}

\begin{document}

\newcommand{\neutralUL}{143}
\newcommand{\chargedUL}{401}
\newcommand{\neutralULdiv}{206}
\newcommand{\chargedULdiv}{577}
\newcommand{\neutralULdivBetter}{159}
\newcommand{\chargedULdivBetter}{446}
\newcommand{\neutralRatio}{3.7}
\newcommand{\chargedRatio}{1.2}
\newcommand{\pippimEff}{2.90\%}
\newcommand{\pipiEff}{1.30\%}
\newcommand{\pimpiEff}{0.660\%}

\title{\texorpdfstring{Upper Limit on the Photoproduction Cross Section of the Spin-Exotic $\pi_1(1600)$}{An Upper Limit on the Photoproduction of the pi1}}

\affiliation{Polytechnic Sciences and Mathematics, School of Applied Sciences and Arts, Arizona State University, Tempe, Arizona 85287, USA}
\affiliation{Department of Physics, National and Kapodistrian University of Athens, 15771 Athens, Greece}
\affiliation{Ruhr-Universit\"{a}t-Bochum, Institut f\"{u}r Experimentalphysik, D-44801 Bochum, Germany}
\affiliation{Helmholtz-Institut f\"{u}r Strahlen- und Kernphysik Universit\"{a}t Bonn, D-53115 Bonn, Germany}
\affiliation{Department of Physics, Carnegie Mellon University, Pittsburgh, Pennsylvania 15213, USA}
\affiliation{Department of Physics, The Catholic University of America, Washington, D.C. 20064, USA}
\affiliation{Department of Physics, University of Connecticut, Storrs, Connecticut 06269, USA}
\affiliation{Department of Physics, Duke University, Durham, North Carolina 27708, USA}
\affiliation{Department of Physics, Florida International University, Miami, Florida 33199, USA}
\affiliation{Department of Physics, Florida State University, Tallahassee, Florida 32306, USA}
\affiliation{Department of Physics, The George Washington University, Washington, D.C. 20052, USA}
\affiliation{School of Physics and Astronomy, University of Glasgow, Glasgow G12 8QQ, United Kingdom}
\affiliation{GSI Helmholtzzentrum f\"{u}r Schwerionenforschung GmbH, D-64291 Darmstadt, Germany}
\affiliation{Institute of High Energy Physics, Beijing 100049, People's Republic of China}
\affiliation{Department of Physics, Indiana University, Bloomington, Indiana 47405, USA}
\affiliation{National Research Centre Kurchatov Institute, Moscow 123182, Russia}
\affiliation{Department of Physics, Lamar University, Beaumont, Texas 77710, USA}
\affiliation{Department of Physics, University of Massachusetts, Amherst, Massachusetts 01003, USA}
\affiliation{National Research Nuclear University Moscow Engineering Physics Institute, Moscow 115409, Russia}
\affiliation{Department of Physics, Mount Allison University, Sackville, New Brunswick E4L 1E6, Canada}
\affiliation{Department of Physics, Norfolk State University, Norfolk, Virginia 23504, USA}
\affiliation{Department of Physics, North Carolina A\&T State University, Greensboro, North Carolina 27411, USA}
\affiliation{Department of Physics and Physical Oceanography, University of North Carolina at Wilmington, Wilmington, North Carolina 28403, USA}
\affiliation{Department of Physics, Old Dominion University, Norfolk, Virginia 23529, USA}
\affiliation{Department of Physics, University of Regina, Regina, Saskatchewan S4S 0A2, Canada}
\affiliation{Department of Mathematics, Physics, and Computer Science, Springfield College, Springfield, Massachusetts, 01109, USA}
\affiliation{Thomas Jefferson National Accelerator Facility, Newport News, Virginia 23606, USA}
\affiliation{Laboratory of Particle Physics, Tomsk Polytechnic University, 634050 Tomsk, Russia}
\affiliation{Department of Physics, Tomsk State University, 634050 Tomsk, Russia}
\affiliation{Department of Physics and Astronomy, Union College, Schenectady, New York 12308, USA}
\affiliation{Department of Physics, Virginia Tech, Blacksburg, VA 24061, USA}
\affiliation{Department of Physics, Washington \& Jefferson College, Washington, Pennsylvania 15301, USA}
\affiliation{Department of Physics, William \& Mary, Williamsburg, Virginia 23185, USA}
\affiliation{School of Physics and Technology, Wuhan University, Wuhan, Hubei 430072, People's Republic of China}
\affiliation{A. I. Alikhanyan National Science Laboratory (Yerevan Physics Institute), 0036 Yerevan, Armenia}
\author{F.~Afzal\orcidlink{0000-0001-8063-6719 }} \affiliation{Helmholtz-Institut f\"{u}r Strahlen- und Kernphysik Universit\"{a}t Bonn, D-53115 Bonn, Germany}
\author{C.~S.~Akondi\orcidlink{0000-0001-6303-5217}} \affiliation{Department of Physics, Florida State University, Tallahassee, Florida 32306, USA}
\author{M.~Albrecht\orcidlink{0000-0001-6180-4297}} \affiliation{Thomas Jefferson National Accelerator Facility, Newport News, Virginia 23606, USA}
\author{M.~Amaryan\orcidlink{0000-0002-5648-0256}} \affiliation{Department of Physics, Old Dominion University, Norfolk, Virginia 23529, USA}
\author{S.~Arrigo} \affiliation{Department of Physics, William \& Mary, Williamsburg, Virginia 23185, USA}
\author{V.~Arroyave} \affiliation{Department of Physics, Florida International University, Miami, Florida 33199, USA}
\author{A.~Asaturyan\orcidlink{0000-0002-8105-913X}} \affiliation{Thomas Jefferson National Accelerator Facility, Newport News, Virginia 23606, USA}
\author{A.~Austregesilo\orcidlink{0000-0002-9291-4429}} \affiliation{Thomas Jefferson National Accelerator Facility, Newport News, Virginia 23606, USA}
\author{Z.~Baldwin\orcidlink{0000-0002-8534-0922}} \affiliation{Department of Physics, Carnegie Mellon University, Pittsburgh, Pennsylvania 15213, USA}
\author{F.~Barbosa} \affiliation{Thomas Jefferson National Accelerator Facility, Newport News, Virginia 23606, USA}
% ...
\collaboration{The \textsc{GlueX} Collaboration}

\date{January 9, 2025}

\begin{abstract}
The spin-exotic hybrid meson $\pi_{1}(1600)$ is predicted to have a large decay rate to the $\omega\pi\pi$ final state. Using 76.6~pb$^{-1}$ of data collected with the GlueX detector, we measure the cross sections for the reactions $\gamma p \to \omega \pi^+ \pi^- p$, $\gamma p \to \omega \pi^0 \pi^0 p$, and $\gamma p\to\omega\pi^-\pi^0\Delta^{++}$ in the range $E_\gamma = 8 - 10$~GeV. Using isospin conservation, we set the first upper limits on the photoproduction cross sections of the $\pi^{0}_{1}(1600)$ and $\pi^{-}_{1}(1600)$. We combine these limits with lattice calculations of decay widths and find that photoproduction of $\eta'\pi$ is the most sensitive two-body system to search for the $\pi_1(1600)$.
\end{abstract}

\maketitle

Quantum chromodynamics (QCD) is the fundamental theory that describes the interactions of quarks and gluons. In nature, protons and neutrons are the most common bound states of QCD, and can be modeled as color-singlet three-quark states called baryons. Even simpler hadronic states are mesons, which consist of just a quark and an antiquark. More complicated configurations are allowed, such as hybrid mesons, in which the gluonic field contributes directly to the total angular momentum $J$, parity $P$, and charge conjugation $C$ quantum numbers of the state. Hybrid mesons are predicted from first principles in lattice QCD (LQCD)~\cite{Dudek:2011tt,Dudek:2013yja}, as well as in phenomenological models of QCD~\cite{Meyer:2015eta,Meyer:2010ku}. Several of the predicted hybrid mesons have spin-exotic quantum numbers, meaning their $J^{PC}$ values are impossible for conventional mesons. LQCD predicts the lightest spin-exotic nonet to have $J^{PC}=1^{-+}$, with the isovector state, the $\pi_1$, being the lightest \cite{Dudek:2013yja}.

Mapping out the spectrum of these hybrid states is a crucial goal of hadronic physics, since it provides important information on the role of gluons in QCD bound states. The best experimental candidate for the lightest hybrid meson is the $\pi_1(1600)$, due to the recent measurements of its decay to $\eta\pi$ and $\eta'\pi$~\cite{COMPASS:2014vkj,JPAC:2018zyd,Kopf:2020yoa}.
Results from LQCD, however, predict the dominant decay mode of the $\pi_1(1600)$ will be $b_1\pi$, with a branching fraction of at least 70\%~\cite{Woss:2020ayi}, which is supported by the fact that there have been several experimental reports of the $\pi_1(1600)$ decaying to $b_{1}\pi$~\cite{VES:1999olf,Amelin:2005ry,E852:2004rfa,Baker:2003jh}. 

While there is considerable experimental evidence for the $\pi_1(1600)$, it has not yet been observed in the photoproduction process. The CLAS experiment has previously studied the charge-exchange reaction $\gamma p\to\pi^+\pi^+\pi^-(n)$ in the range $E_\gamma = 4.8 - 5.4$~GeV, and found that the dominant contributions were from the conventional $a_2^+(1320)$ and $\pi_2^+(1670)$ mesons, with no evidence for the spin-exotic $\pi_1^+(1600)$ decay to $\rho\pi$~\cite{CLAS:2008zko}. Given the large predicted decay of the $\pi_1(1600)$ to $b_{1}\pi$ and the dominant decay of the $b_{1}$ to $\omega\pi$~\cite{ParticleDataGroup:2022pth,Woss:2019hse}, in this paper we aim to study the isospin-1 contribution to $\omega\pi\pi$ photoproduction. The existing photoproduction results for $\omega\pi^{+}\pi^{-}$ suggest the isospin-0 contribution is larger than the isospin-1 contribution~\cite{Blackett:1997cm,OmegaPhoton:1983eln,OmegaPhoton:1986aqd}, which implies it will be difficult to study the isospin-1 component using only $\omega\pi^+\pi^-$ data. However, with access to different charge combinations of the $\omega\pi\pi$ system and by exploiting isospin symmetry, it is possible to separate the isospin-0 and isospin-1 contributions. There are no known isospin-2 (i.e., flavor-exotic) states, so we assume no isospin-2 contributions to the $\omega\pi\pi$ distributions. Under this assumption, the $\omega\pi^{0}\pi^{0}$ system is purely isospin-0, and the $\omega\pi^{+}\pi^{-}$ system contains both isospin-0 and isospin-1. Using isospin Clebsch-Gordan coefficients, we find the isospin-0 cross section
\begin{eqnarray}
    \label{eq:i=0}
    \sigma[\omega\pi\pi]_{I=0} &= & 3\sigma[\omega\pi^0\pi^0].
\end{eqnarray}
The total neutral $\omega\pi\pi$ cross section can be written as $\sigma[\omega\pi\pi]_{I=0}+\sigma[\omega\pi\pi]_{I=1}=\sigma[\omega\pi^+\pi^-]+\sigma[\omega\pi^0\pi^0]$, so solving for $\sigma[\omega\pi\pi]_{I=1}$ gives the isospin-1 cross section
\begin{eqnarray}
 \label{eq:i=1}
    \sigma[\omega\pi\pi]_{I=1} & = & \sigma[\omega\pi^+\pi^-]-2\sigma[\omega\pi^0\pi^0].
\end{eqnarray}

In this Letter, we measure the cross sections for
\begin{Reqnarray}
\label{Req:reac1}
\gamma p & \rightarrow & \omega\pi^{0}\pi^{0}p \\
\label{Req:reac2}
\gamma p & \rightarrow & \omega \pi^{+}\pi^{-}p \\
\label{Req:reac3}
\gamma p & \rightarrow & \omega\pi^{-}\pi^{0}\Delta^{++} 
\end{Reqnarray}
where we reconstruct $\omega\to\pi^+\pi^-\pi^0$, $\pi^0\to\gamma\gamma$, and $\Delta^{++}\to \pi^+p$. We isolate the isospin-1 $\omega\pi\pi$ cross section produced against a proton using Eq.~\ref{eq:i=1} with \ref{Req:reac1} and \ref{Req:reac2}, while \ref{Req:reac3} gives the isospin-1 $\omega\pi\pi$ cross section with a recoil $\Delta^{++}$. We observe no clear structure in the mass spectrum that is consistent with a Breit-Wigner amplitude for the $\pi_1(1600)$. A comprehensive search for $\pi_1(1600)\rightarrow\omega\pi\pi$ would require a multidimensional angular analysis, which is beyond the scope of this work. However, we can use the isospin-1 $\omega\pi\pi$ spectra to set a meaningful upper limit on the $\pi_1(1600)$ photoproduction cross section, which can be used to guide our study of $\eta\pi$ and $\eta'\pi$ production. Using known resonance parameters for the $a_2(1320)$ and $\pi_1(1600)$, we determine the largest $\pi_1(1600)\to b_1\pi\to \omega\pi\pi$ contribution that is consistent with our measured $\omega\pi\pi$ data. These limits are combined with the allowed ranges of decay widths for $\pi_1(1600)\to\eta\pi$ and $\pi_1(1600)\to\eta'\pi$ from Ref.~\cite{Woss:2020ayi} to estimate the maximal $\pi_1(1600)$ contributions to the photoproduced $\eta^{(\prime)}\pi$ mass spectra. In particular, we will show that the $\eta\pi$ final states can have at most a percent-level contribution from the $\pi_1(1600)$, while our limit cannot rule out a dominant $\pi_1(1600)$ contribution to the $\eta'\pi$ systems. This leads us to the conclusion that in photoproduction, the $\eta'\pi$ final states are the most sensitive two-body systems to search for the $\pi_1(1600)$.

The GlueX experiment at Jefferson Lab has acquired a world-leading photoproduction data sample, with a goal of mapping out the hybrid meson spectrum. The linearly polarized photons are produced via the coherent bremsstrahlung process by an electron beam incident on a $50~$\textmu m thick diamond wafer, where the maximum photon polarization of about 35\% occurs in the coherent peak at a photon energy of $8.8\,\text{GeV}$. Events with photon energies outside the coherent peak are also recorded, leading to a tagged photon energy spectrum between $7.0\,\text{GeV}$ and $11.6\,\text{GeV}$. The photon beam is then incident on a liquid hydrogen target. GlueX is a nearly hermetic detector that reconstructs charged particles and photons. Particle identification is available from both $dE/dx$ in the drift chambers and flight-time measurements in the central calorimeter and a dedicated downstream detector. Detailed information on the beamline and detector can be found elsewhere~\cite{GlueX:2020idb,Barbosa:2015bga,Berdnikov:2015jja,Dugger:2017zoq,Pentchev:2017omk,Beattie:2018xsk,Pooser:2019rhu,Jarvis:2019mgr}.

Events with photon beam energies between $8.0$~GeV and 10.0~GeV are selected from about 1/3 of the Phase-I GlueX data (corresponding to one running period), yielding a total luminosity of $76.6$~pb$^{-1}$. We fully reconstruct events with the final states for reactions~\ref{Req:reac1},~\ref{Req:reac2}, or~\ref{Req:reac3}. We keep events with ${0.1\,(\text{GeV}/c)^2<-t<0.5\,(\text{GeV}/c)^{2}}$, where $-t$ is the four-momentum transfer squared to the recoil baryon. A kinematic fit to enforce energy and momentum conservation with mass constraints on $\pi^{0}\rightarrow\gamma\gamma$ is then performed, and events with a resulting $\chi^{2}/\mathrm{ndf}$ smaller than $5$ are retained for further analysis. Beam bunches arrive every 4~ns, so we select beam photon candidates within $\pm$2~ns of the beam bunch that corresponds to the reconstructed final state particles. For each event, there can be multiple beam photon candidates, as well as multiple ways to assign photons to the $\pi^0$ candidates. To select a unique combination for a given event, we use only the beam photon and final-state particle combination with the smallest $\chi^{2}/\mathrm{ndf}$. For~\ref{Req:reac3}, we require the invariant mass $M(\pi^+p)<1.35\,\text{GeV}/c^2$ to select the $\Delta^{++}$. For~\ref{Req:reac2}, background studies show a large contamination by $\gamma p\to b_1^-\Delta^{++}$, which we remove by requiring $M(\pi^+p)>1.35$~GeV/$c^2$ for both $\pi^+$ combinations in the event. After this requirement, the remaining baryon backgrounds are small for all three reactions. Background studies also show a large background for~\ref{Req:reac1} from $\gamma p \to \omega\pi^0\pi^0\pi^0 p$, which we suppress by requiring that no additional reconstructed calorimeter showers are present in the event. Background Monte Carlo studies show no more than 3\% of the remaining $\omega\pi^0\pi^0$ events are due to the $\omega\pi^0\pi^0\pi^0$ background.

Monte Carlo (MC) samples are used to estimate the reconstruction efficiency for the $\omega\pi\pi$ events. The MC samples are generated by modeling the decay $\pi_1(1600)\to\omega\pi\pi$ as a three-body decay with a uniform distribution in phase space. We assume an exponential distribution for the four-momentum transfer squared $-t$, given by $e^{bt}$, with $b=5$~(GeV/$c$)$^{-2}$ chosen to approximate the $-t$ distribution in data. The $\pi_1(1600)$ shape is modeled as a relativistic Breit-Wigner, with the mass and width taken as the pole parameters from Ref.~\cite{JPAC:2018zyd}, since it is the only analysis that simultaneously describes the $\eta\pi$ and $\eta'\pi$ data. This simulated data is passed through a GEANT4~\cite{geant4} model of the detector, and is analyzed using the same selection criteria as for the data. To determine the reconstruction efficiency, we perform a binned maximum likelihood fit on the reconstructed $\omega$ candidate masses in signal MC with a Voigtian~\cite{Batty:1976cm}, where the natural width of the $\omega$ is fixed to the value from the Particle Data Group~\cite{ParticleDataGroup:2022pth}. This leaves the $\omega$ mass, the resolution, and the normalization as floating parameters in the fit to MC data. We integrate the resulting Voigtian and divide by the number of generated events to determine our reconstruction efficiency. We then fix the mass and resolution parameters of the Voigtian to the results of the MC fits, leaving the overall normalization as the only free parameter in fits to data.

To measure the differential $\omega\pi\pi$ photoproduction cross sections $d\sigma/dm_{\omega\pi\pi}$, we divide the data into $20$ equidistant bins of $\omega\pi\pi$ invariant mass $m_{\omega\pi\pi}$ from $1.20$~GeV/$c^2$ to $2.20$~GeV/$c^{2}$. For each of these bins, we plot the invariant mass for each $\pi^{+}\pi^{-}\pi^{0}$ combination. We fit these data using a least-$\chi^2$ fit with a Voigtian function for the $\omega$ signal plus a fourth-order polynomial for the background. These fits are then used to determine the $\omega\pi\pi$ yield $Y_{\omega\pi\pi}$ for each bin. Sample fits to the $M(\pi^+\pi^-\pi^0)$ distributions are shown in Sec.~I of the Supplemental Material~\cite{SuppMaterials}.

\begin{figure*}[htbp!] \centering
\begin{overpic}[scale=0.27]{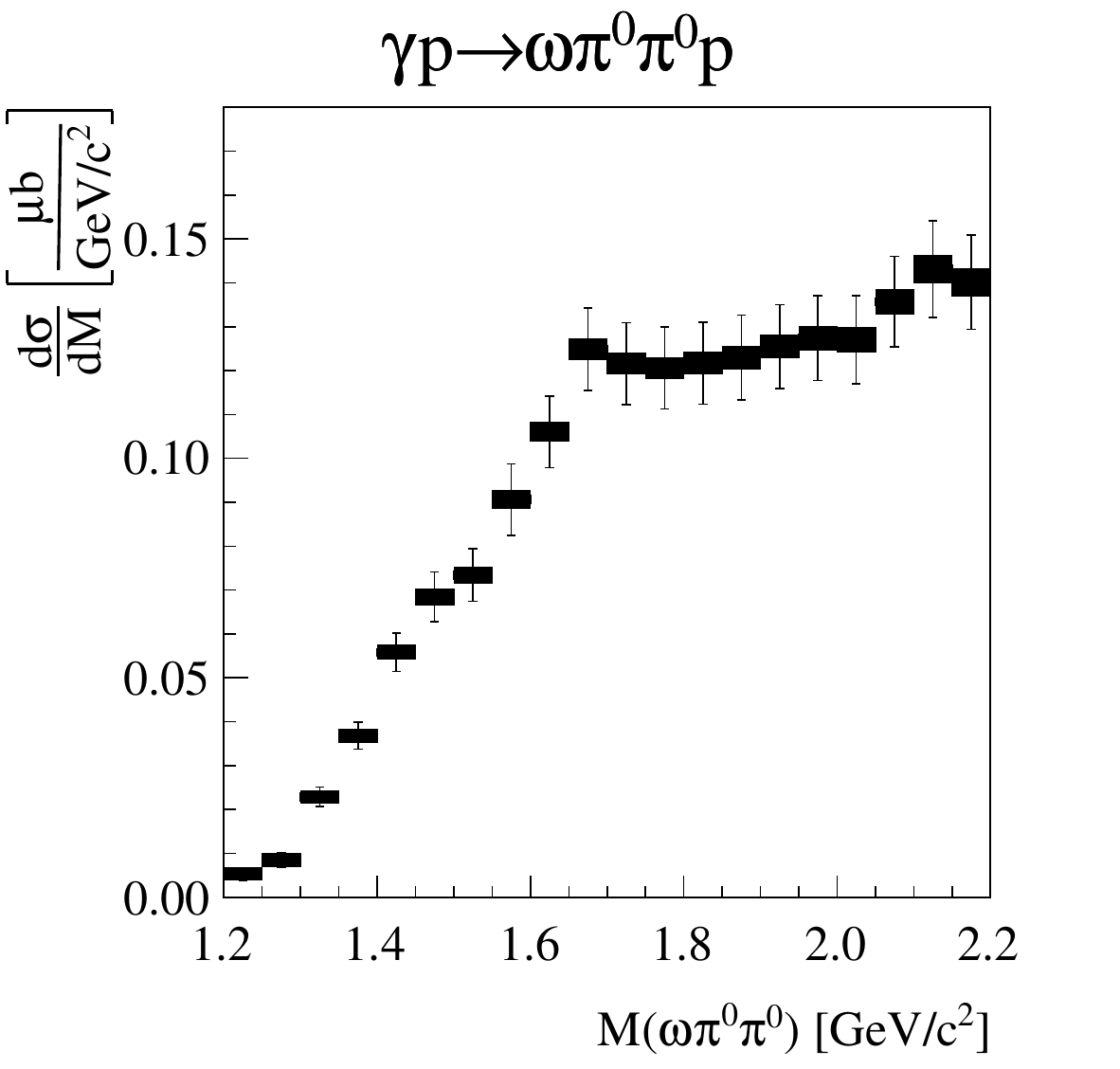}
\put(23,75){(a)}
\put(37,25){Purely $I=0$}
\end{overpic}\begin{overpic}[scale=0.27]{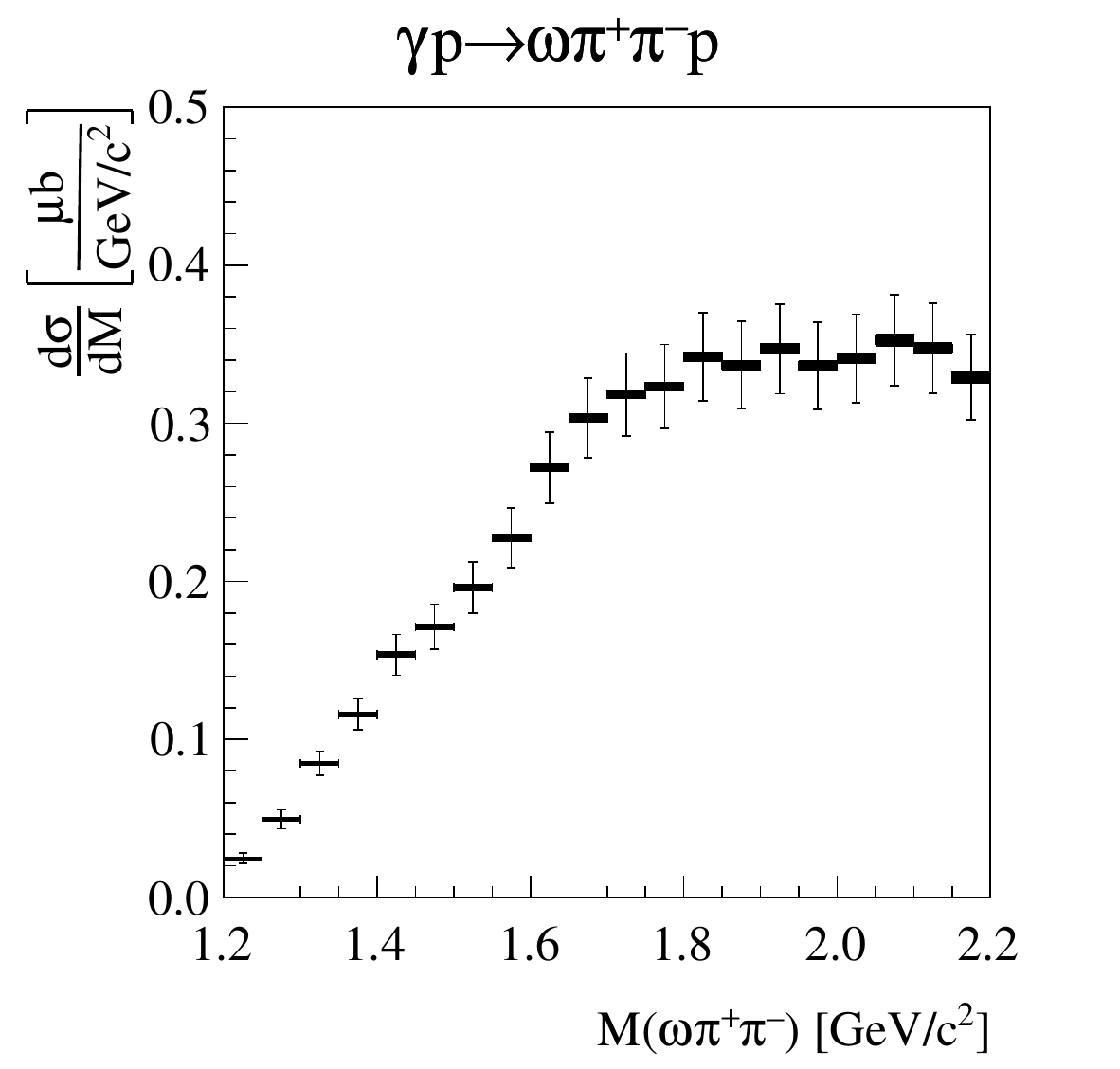}
\put(23,75){(b)}
\put(37,25){$I=1$ and $I=0$}
\end{overpic}\begin{overpic}[scale=0.27]{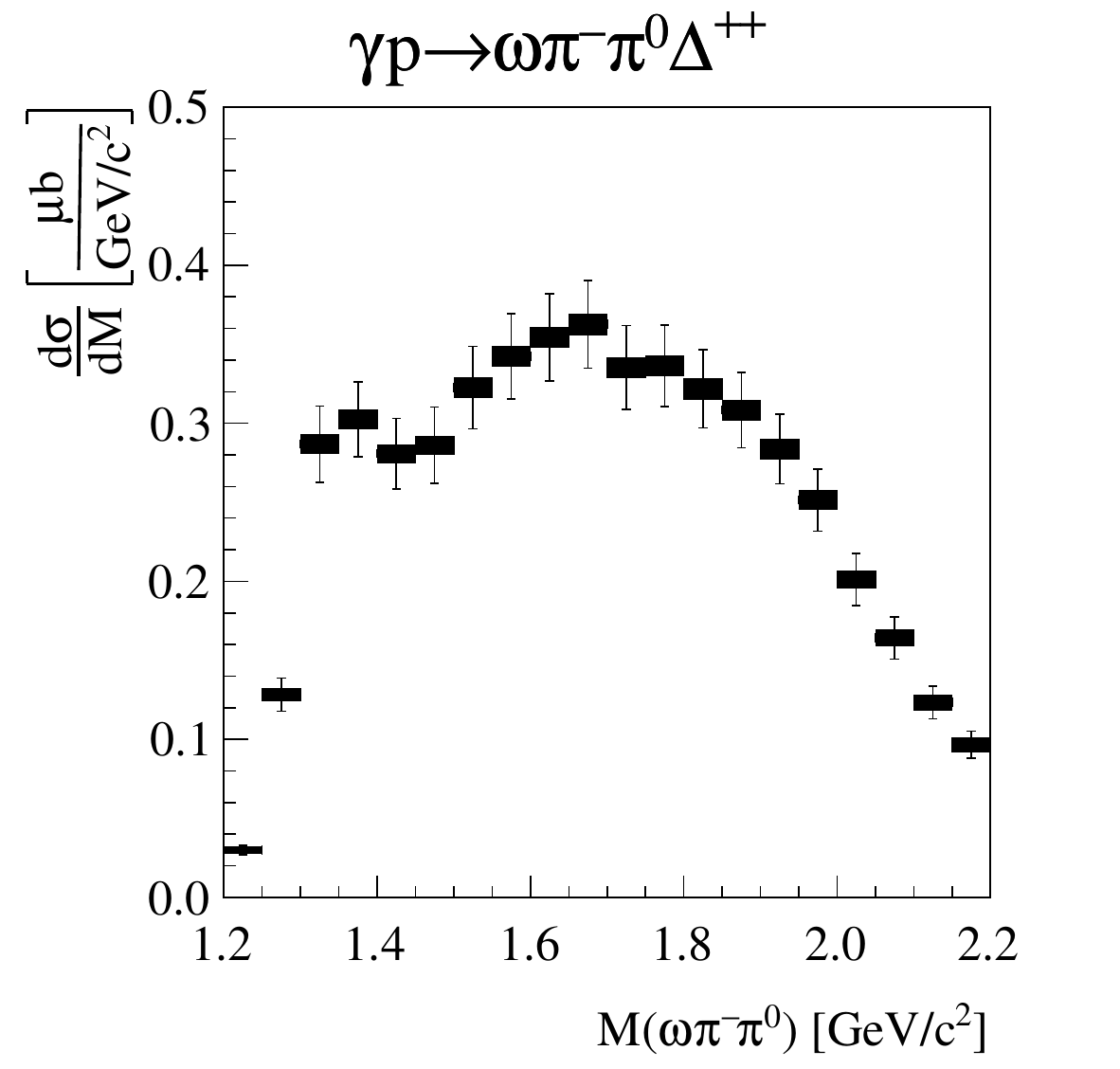}
\put(23,75){(c)}
\put(37,25){Purely $I=1$}
\end{overpic}
    \caption{The differential cross sections for the reactions $\gamma p \to \omega \pi^0\pi^0 p$ (a), $\gamma p \to \omega \pi^+\pi^- p$ (b), and  $\gamma p \to \omega\pi^-\pi^0\Delta^{++}$ (c) with $0.1\,(\text{GeV}/c)^2<\, -t \,< 0.5\,(\text{GeV}/c)^2$ as a function of $\omega\pi\pi$ invariant mass. The filled rectangles show the statistical uncertainty, and the full error bars are statistical and systematic uncertainties added in quadrature. Not included in the error bars are uncertainties due to the photon and charged-track efficiency systematic uncertainty, as well as the systematic uncertainty due to the luminosity. These uncertainties are correlated for each source across the three measurements, so they cannot be easily visualized.}
    \label{fig:xs}
\end{figure*}

We use the yields and reconstruction efficiencies to determine the cross sections for \ref{Req:reac1}$-$\ref{Req:reac3} in each bin of $\omega\pi\pi$ mass. The measured cross sections are shown in Fig.~\ref{fig:xs}. We expect the $a_2(1320)$ to contribute to both the $\gamma p \to \omega\pi^+\pi^- p$ and $\gamma p\to \omega\pi^-\pi^0\Delta^{++}$ reactions, but only the charge-exchange reaction shows a clear $a_2(1320)$ peak. This is due to the fact that the $\omega\pi^+\pi^-$ system has both isospin-1 and isospin-0 components, while the $\omega\pi^-\pi^0$ system is purely isospin-1. This implies that the $\omega\pi^+\pi^-$ mass spectrum is dominated by isospin-0, which is consistent with the previous results from the literature~\cite{OmegaPhoton:1983eln}. 

\begin{figure*}[htbp!]\centering
\begin{overpic}[scale=0.28,page=5]{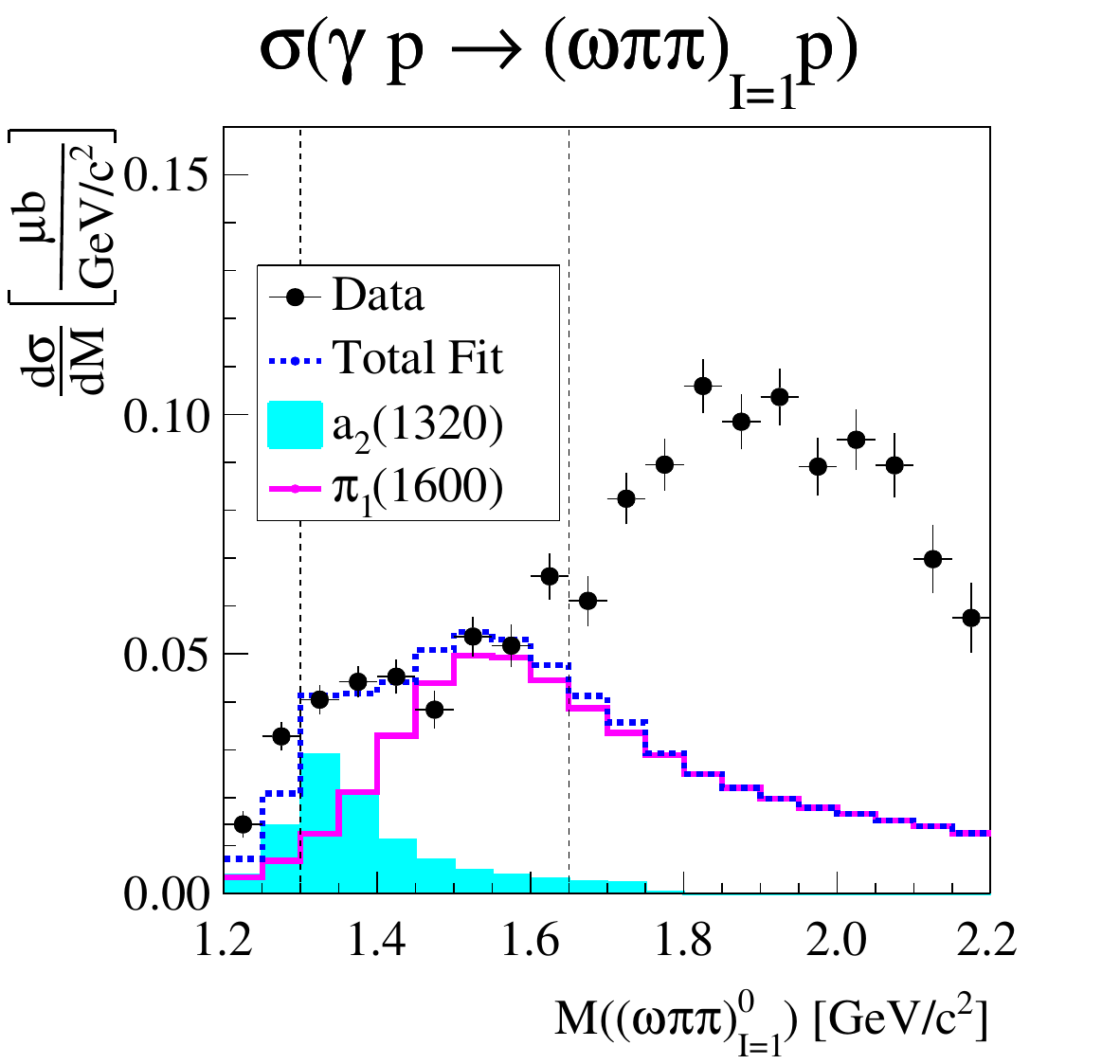}
\put(23,75){(a)}
\end{overpic}\begin{overpic}[scale=0.28,page=5]{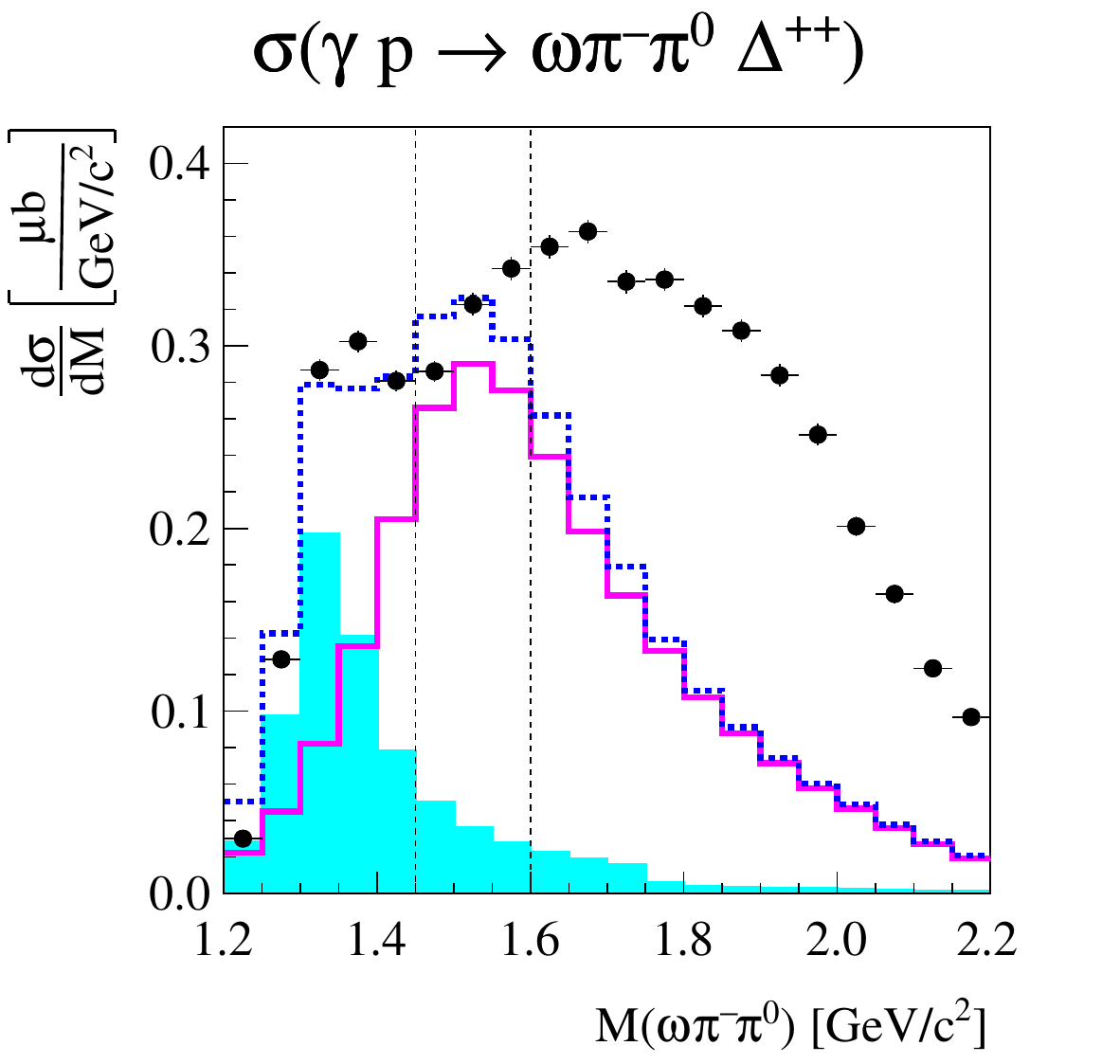}
\put(23,75){(b)}
\end{overpic}
    \caption{The isospin-1 component of the cross sections for neutral (a) and negatively charged (b) $\omega\pi\pi$ for $0.1\,(\text{GeV}/c)^2<\, -t \,< 0.5\,(\text{GeV}/c)^{2}$. The error bars are statistical only. The cross sections are fit with an $a_2(1320)$ (cyan) and $\pi_1(1600)$ (magenta) Breit-Wigner shape in the range between the vertical dashed lines.}
    \label{fig:sep}
\end{figure*}

As a next step, we use \ref{Req:reac1}$-$\ref{Req:reac3} and Eq. \ref{eq:i=1} to determine the cross sections for producing an isospin-1 $\omega\pi\pi$ system recoiling against a proton or a $\Delta^{++}$, which we show in Fig.~\ref{fig:sep}. We could have contributions from nonresonant $\omega\pi\pi$, the $a_2(1320)$, the $\pi_2(1670)$, the $a_2(1700)$, and the $\pi_1(1600)$. Previous studies of $\gamma p\to \rho^0\pi^- \Delta^{++}$ at the SLAC Hybrid Facility show a large $a_2(1320)$ signal, and a substantially smaller signal near the $\pi_2(1670)$ mass \cite{CONDO:SLAC1993}. This, coupled with the fact that the $\pi_2(1670)\to\omega\rho$ branching fraction is just 2.7\% while $\mathcal{B}[\pi_2(1670)\to3\pi]=95.8\%$ \cite{ParticleDataGroup:2022pth}, implies that the $\pi_2(1670)$ should be a small contribution to $\omega\pi\pi$. The $a_2(1700)$ cross section has not been measured in photoproduction, but assuming that both the cross section and the branching fraction to $\omega\pi\pi$ are similar to those for the $a_2(1320)$, the $a_2(1700)$ would make up a small fraction of the events in the $\pi_1(1600)$ mass region.

We perform a binned least-$\chi^2$ fit to these isospin-1 $\omega\pi\pi$ photoproduction cross sections with the sum of an $a_2(1320)$ and $\pi_1(1600)$ shape to determine the largest $\pi_1(1600)$ signal consistent with the data. The $a_2(1320)$ shape is a fixed-width relativistic Breit-Wigner with $M=1318.2$~MeV/$c^2$ and $\Gamma=110$~MeV \cite{ParticleDataGroup:2022pth}, while the $\pi_1(1600)$ shape is a fixed-width relativistic Breit-Wigner using the Joint Physics Analysis Center parameters \cite{JPAC:2018zyd} from their fit to the COMPASS data \cite{COMPASS:2014vkj}, i.e., $M_{\pi_1(1600)}=(1564\pm 24\pm 86)$~MeV$/c^2$ and $\Gamma_{\pi_1(1600)}=(492\pm 54\pm 102)$~MeV. The magnitude of the $a_2^0(1320)$ cross section is fixed based on the results of a partial-wave analysis to $\gamma p \to \eta\pi^0 p$ \cite{Malte:Hadron} with $\eta\to\gamma\gamma$, while the magnitude of the $a_2^-(1320)$ cross section is fixed based on the results of a one-dimensional fit to the $M(\eta\pi^-)$ system in the reaction $\gamma p \to \eta\pi^-\Delta^{++}$ with $\eta\to\pi^+\pi^-\pi^0$, as discussed in Ref.~\cite{SuppMaterials} Sec.~II. We use the $\eta\pi$ systems to fix the size of the $a_2(1320)$ cross sections because the $a_2(1320)$ is more significant in those reactions. The only free parameters in the fits to the isospin-1 cross sections are the overall magnitude of the $\pi_1^0(1600)$ or $\pi_1^-(1600)$. Since there are no other floating contributions in the fits, we only fit the region around the $\pi_1(1600)$ peak to find the largest $\pi_1(1600)$ signal that does not exceed the data. Note that the fit model cannot precisely match the shape of the data since no background contributions are included.  We include a discussion of the systematic uncertainties in Ref.~\cite{SuppMaterials} Secs.~IV$-$VI.

In order to determine the upper limits on the $\pi_1(1600)$ cross sections, we need to determine how the likelihood varies as a function of $\sigma[\gamma p\to\pi_1^0(1600)p]$ and $\sigma[\gamma p\to \pi_1^-(1600)\Delta^{++}]$, which we denote $\sigma_x$ for simplicity. The likelihood function is given by $e^{[\chi^2_{\text{min}}-\chi^2(\sigma_x)]/2}$ where $\chi^2_{\text{min}}$ is the $\chi^2$ of the best fit compared to the data and $\chi^2(\sigma_x)$ is the $\chi^2$ for an alternative value of $\sigma_x$. The 90\% confidence level upper limit is given by the 90th percentile of this likelihood distribution, after accounting for the systematic uncertainties as described in Ref.~\cite{SuppMaterials} Sec.~VI.

 After including the systematic uncertainties, we place 90\% confidence level (CL) upper limits of 
 \begin{equation}
     \sigma[\gamma p\to \pi_1^0(1600) p]\times\mathcal{B}[\pi_1(1600)\to b_1\pi]<\neutralUL~\text{nb}
 \end{equation} and 
 \begin{equation}
     \sigma[\gamma p\to \pi_1^-(1600)\Delta^{++}]\times\mathcal{B}[\pi_1(1600)\to b_1\pi]<\chargedUL~\text{nb}
 \end{equation} where we assume $\mathcal{B}[b_1\to\omega\pi]=100\%$.  Several systematic uncertainties cancel when measuring a ratio of cross sections, including all of the luminosity systematic uncertainty and portions of the photon and tracking efficiency uncertainties. Therefore, we also report our limits normalized to the measured $a_2(1320)$ cross sections. After accounting for these reduced systematic uncertainties, as well as incorporating the uncertainties on the measured $a_2(1320)$ cross sections as discussed in Ref.~\cite{SuppMaterials} Sec.~VII, we place 90\% CL upper limits of 
\begin{equation} \frac{\sigma[\gamma p\to\pi_1^0(1600) p]\times\mathcal{B}[\pi_1(1600)\to b_1\pi]}{\sigma[\gamma p\to a_2^0(1320) p]} < \neutralRatio
\end{equation}
and
\begin{equation} \frac{\sigma[\gamma p\to\pi_1^-(1600)\Delta^{++}]\times\mathcal{B}[\pi_1(1600)\to b_1\pi]}{\sigma[\gamma p\to a_2^-(1320) \Delta^{++}]} < \chargedRatio.
\end{equation}
 
Lattice calculations from Ref.~\cite{Woss:2020ayi} include a range of allowed values for $\mathcal{B}[\pi_1(1600)\to b_1\pi]$. We can use the lower limit on $\mathcal{B}[\pi_1(1600)\to b_1\pi]$ from their result to set an upper limit directly on the $\pi_1(1600)$ photoproduction cross section, with the caveat that it is no longer a 90\% CL limit since the uncertainty on $\mathcal{B}[\pi_1(1600)\to b_1\pi]$ is not folded in. Using their lower bound on $\mathcal{B}[\pi_1(1600)\to b_1\pi]$, we set limits of $\sigma[\gamma p\to \pi_1^0(1600)p]<\neutralULdiv$~nb and $\sigma[\gamma p \to \pi_1^-(1600)\Delta^{++}]<\chargedULdiv$~nb. These are the first upper limits on the $\pi_1^0(1600)$ and $\pi_1^-(1600)$ photoproduction cross sections. Note that no other reaction can be used to set an upper limit on the $\pi_1(1600)$ since $b_1\pi$ is the only decay mode that has a calculated lower limit.

As mentioned above, these limits on the $\pi_1(1600)$ photoproduction cross section can be used with the calculated $\pi_1(1600)$ decay widths from LQCD~\cite{Woss:2020ayi} to determine which two-body channels are most sensitive to the $\pi_1(1600)$ in photoproduction. To do this, we need two pieces of information: the maximum size of the $\pi_1(1600)$ signal in the channel and the total size of the photoproduction cross section for the channel. To estimate the maximum $\pi_1(1600)$ contribution in the $\eta^{(\prime)}\pi$ channels, we assume $\mathcal{B}[b_1\to\omega\pi]=100\%$ and multiply the measured upper limit on $\sigma[\gamma p \to \pi_1^0(1600)p]\times\mathcal{B}[\pi_1(1600)\to b_1\pi]$ by $R=\mathcal{B}[\pi_1(1600)\to \eta^{(\prime)}\pi]/\mathcal{B}[\pi_1(1600)\to b_1\pi]$. The maximum $\pi_1(1600)$ contribution comes when $R$ is largest, so we maximize $\Gamma[\pi_1(1600)\to\eta^{(\prime)}\pi]$ and minimize $\Gamma[\pi_1(1600)\to b_1\pi]$ under the constraints that $\Gamma_{\text{tot}}(\pi_1(1600))=492$~MeV and that the partial widths are in the ranges allowed by Ref.~\cite{Woss:2020ayi}.  The corresponding values are $R=2.8\times 10^{-2}$ for $\eta'\pi$ and $R=2.3\times 10^{-3}$ for $\eta\pi$. More details on how these values were obtained are in Ref.~\cite{SuppMaterials} Sec.~III.

\begin{figure*}
    \centering
    \includegraphics[scale=0.55]{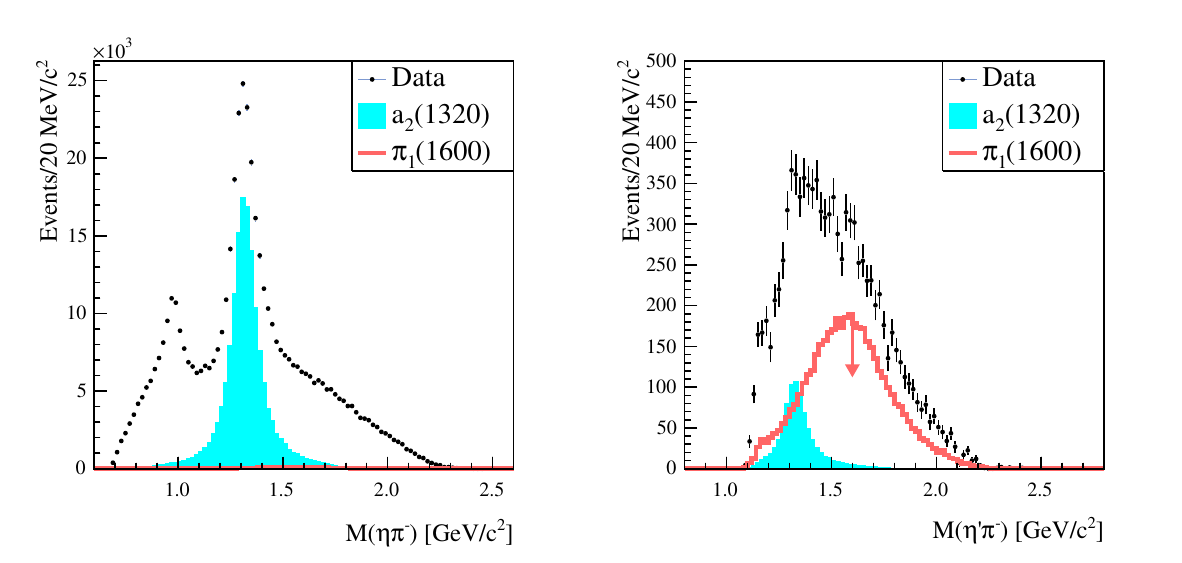}
    \caption{The reconstructed $\eta^{(\prime)}\pi^-$ invariant mass distributions (points), overlaid with the $a_2(1320)$ signal (cyan) and the $\pi_1(1600)$ upper limit (red). }
    \label{fig:projections}
\end{figure*}

Next, we compare these limits to the total size of the $\eta\pi$ and $\eta'\pi$ photoproduction cross sections. We select exclusive events for the processes $\gamma p\to \eta^{(\prime)}\pi^0p$ and $\gamma p\to \eta^{(\prime)}\pi^- \Delta^{++}$ with $\Delta^{++}\to\pi^+p$, $\eta'\to\pi^+\pi^-\eta$, and $\eta\to\gamma\gamma$. We describe the selection criteria for these four reactions in Ref.~\cite{SuppMaterials}, Sec.~III. The reconstructed $\eta^{(\prime)}\pi^-$ invariant mass distributions are shown in Fig.~\ref{fig:projections}, and the corresponding plots for $\eta^{(\prime)}\pi^0$ are in Ref.~\cite{SuppMaterials}, Sec.~III. Note that we are showing the number of reconstructed events rather than the cross section, so we convert our $a_2(1320)$ cross section and $\pi_1(1600)$ upper limit to an expected number of events using the equation $N=\sigma \mathcal{L}_{\text{int}}\epsilon \mathcal{B}$, where $\sigma$ is the cross section, $\mathcal{L}_{\text{int}}$ is the integrated luminosity, $\epsilon$ is the reconstruction efficiency, and $\mathcal{B}$ is the product of all required branching fractions. The figure shows the measured size of the $a_2(1320)$ signal, as well as the projected $\pi_1(1600)$ upper limit. These projections show that the $\pi_1(1600)$ will be less than 1\% of the total $\eta\pi$ mass spectra, but it could be the main contribution to the $\eta'\pi$ mass spectra. 

In conclusion, we have measured the cross sections for the photoproduction of three different $\omega\pi\pi$ final states off of a proton target, and used them to set an upper limit for the photoproduction cross section of the lightest hybrid meson candidate. We find that the upper limits on the $\pi_1(1600)$ cross sections are similar in size to the $a_{2}(1320)$ cross section. Assuming $\mathcal{B}[b_1\to\omega\pi]=100\%$, we place 90\% CL upper limits of  $\sigma[\gamma p\to \pi_1^0(1600) p]\times\mathcal{B}[\pi_1(1600)\to b_1\pi]<\neutralUL~\text{nb}$ and $\sigma[\gamma p\to \pi_1^-(1600)\Delta^{++}]\times\mathcal{B}[\pi_1(1600)\to b_1\pi]<\chargedUL~\text{nb}$. We combine these results with lattice QCD calculations for the $\pi_1(1600)$ decay widths and find $\sigma[\gamma p \to \pi_1^0(1600) p]<\neutralULdiv$~nb and $\sigma[\gamma p \to \pi_1^-(1600)\Delta^{++}]<\chargedULdiv$~nb for $0.1\,(\text{GeV}/c)^2<-t<0.5\,(\text{GeV}/c)^2$. These are the first limits on $\pi_1(1600)$ photoproduction, since $\pi_1\to b_1\pi$ is the only decay that has a calculated lower limit on the size of its branching fraction. We also find that the best discovery potential for the $\pi_1(1600)$ in photoproduction is in the $\eta'\pi$ decay modes, where the $\pi_1(1600)$ could be a dominant contribution, and where it would appear as an exotic $P$-wave amplitude, which is forbidden for conventional mesons. The $\pi_1(1600)$ could also be dominant in the $\gamma p\to \omega\pi^-\pi^0\Delta^{++}$ reaction, but the dominantly $S$-wave decay of $\pi_1(1600)\to b_1\pi$ combined with the broadness of the $b_1$ makes this analysis much more difficult.

\section{Acknowledgements}
The analysis in this article was supported by the U.S. Department of Energy, Office of Science, Office of Nuclear Physics under contract DOE Grant No. DE-FG02-87ER40315. We would like to acknowledge the outstanding efforts of the staff of the Accelerator and the Physics Divisions at Jefferson Lab that made the experiment possible. This work was also supported in part by the U.S. Department of Energy, the U.S. National Science Foundation, NSERC Canada, the German Research Foundation, GSI Helmholtzzentrum f\"{u}r Schwerionenforschung GmbH, the Russian Foundation for Basic Research, the UK Science and Technology Facilities Council, the Chilean Comisi\'{o}n Nacional de Investigaci\'{o}n Cient\'{i}fica y Tecnol\'{o}gica, the National Natural Science Foundation of China, and the China Scholarship Council. This material is based upon work supported by the U.S. Department of Energy, Office of Science, Office of Nuclear Physics under Contract No. DE-AC05-06OR23177. This research used resources of the National Energy Research Scientific Computing Center (NERSC), a U.S. Department of Energy Office of Science User Facility operated under Contract No. DE-AC02-05CH11231. This work used the Extreme Science and Engineering Discovery Environment (XSEDE), which is supported by National Science Foundation Grant No. ACI-1548562. Specifically, it used the Bridges system, which is supported by NSF Award No. ACI-1445606, at the Pittsburgh Supercomputing Center (PSC).

\begin{thebibliography}{0}%
\makeatletter
\providecommand \@ifxundefined [1]{%
 \@ifx{#1\undefined}
}%
\providecommand \@ifnum [1]{%
 \ifnum #1\expandafter \@firstoftwo
 \else \expandafter \@secondoftwo
 \fi
}%
\providecommand \@ifx [1]{%
 \ifx #1\expandafter \@firstoftwo
 \else \expandafter \@secondoftwo
 \fi
}%
\providecommand \natexlab [1]{#1}%
\providecommand \enquote  [1]{``#1''}%
\providecommand \bibnamefont  [1]{#1}%
\providecommand \bibfnamefont [1]{#1}%
\providecommand \citenamefont [1]{#1}%
\providecommand \href@noop [0]{\@secondoftwo}%
\providecommand \href [0]{\begingroup \@sanitize@url \@href}%
\providecommand \@href[1]{\@@startlink{#1}\@@href}%
\providecommand \@@href[1]{\endgroup#1\@@endlink}%
\providecommand \@sanitize@url [0]{\catcode `\\12\catcode `\$12\catcode
  `\&12\catcode `\#12\catcode `\^12\catcode `\_12\catcode `\%12\relax}%
\providecommand \@@startlink[1]{}%
\providecommand \@@endlink[0]{}%
\providecommand \url  [0]{\begingroup\@sanitize@url \@url }%
\providecommand \@url [1]{\endgroup\@href {#1}{\urlprefix }}%
\providecommand \urlprefix  [0]{URL }%
\providecommand \Eprint [0]{\href }%
\providecommand \doibase [0]{https://doi.org/}%
\providecommand \selectlanguage [0]{\@gobble}%
\providecommand \bibinfo  [0]{\@secondoftwo}%
\providecommand \bibfield  [0]{\@secondoftwo}%
\providecommand \translation [1]{[#1]}%
\providecommand \BibitemOpen [0]{}%
\providecommand \bibitemStop [0]{}%
\providecommand \bibitemNoStop [0]{.\EOS\space}%
\providecommand \EOS [0]{\spacefactor3000\relax}%
\providecommand \BibitemShut  [1]{\csname bibitem#1\endcsname}%
\let\auto@bib@innerbib\@empty
\end{thebibliography}%


\begin{thebibliography}{9}

\bibitem{Dudek:2011tt}
    J. J. Dudek, R. G. Edwards, B. Jo\'o, M. J. Peardon, D. G. Richards, and C. E. Thomas (Hadron Spectrum Collaboration), \href{https://doi.org/10.1103/PhysRevD.83.111502}{Phys. Rev. D \textbf{83}, 111502(R) (2011).}

\bibitem{Dudek:2013yja}
    J. J. Dudek, R. G. Edwards, P. Guo, and C. E. Thomas (Hadron Spectrum Collaboration), \href{https://doi.org/10.1103/PhysRevD.88.094505}{Phys. Rev. D \textbf{88}, 094505 (2013).}

\bibitem{Meyer:2015eta}
    C. A. Meyer and E. S. Swanson, \href{https://doi.org/10.1016/j.ppnp.2015.03.001}{Prog. Part. Nucl. Phys. \textbf{82}, 21 (2015).}

\bibitem{Meyer:2010ku}
    C. A. Meyer and Y. Van Haarlem, \href{https://doi.org/10.1103/PhysRevC.82.025208}{Phys. Rev. C \textbf{82}, 025208 (2010).}

\bibitem{COMPASS:2014vkj}
    C. Adolph \textit{et al.} (COMPASS Collaboration) \href{https://doi.org/10.1016/j.physletb.2014.11.058}{Phys. Lett. B \textbf{740}, 303 (2015)};\href{https://doi.org/10.1016/j.physletb.2020.135913}{\textbf{811}, 135913(R) (2020).}

\bibitem{JPAC:2018zyd}
    A. Rodas \textit{et al.} (Joint Physics Analysis Center), \href{https://doi.org/10.1103/PhysRevLett.122.042002}{Phys. Rev. Lett. \textbf{122}, 042002 (2019).}

\bibitem{Kopf:2020yoa}
    B.~Kopf, M.~Albrecht, H.~Koch, M.~K\"u\ss{}ner, J.~Pychy, X.~Qin and U.~Wiedner,
    \href{https://doi.org/10.1140/epjc/s10052-021-09821-2}{Eur. Phys. J. C \textbf{81}, 1056 (2021).}


\bibitem{Woss:2020ayi}
    A. J. Woss, J. J. Dudek, R. G. Edwards, C. E. Thomas, and D. J. Wilson (Hadron Spectrum Collaboration), 
\href{https://doi.org/10.1103/PhysRevD.103.054502}{Phys. Rev. D \textbf{103}, 054502 (2021).}

\bibitem{VES:1999olf}
    D. V. Amelin \textit{et al.} (VES Collaboration), Phys. At. Nucl. \textbf{62}, 445 (1999).

\bibitem{Amelin:2005ry}
    D. V. Amelin \textit{et al.}, \href{https://doi.org/10.1134/1.1891185}{Phys. At. Nucl. \textbf{68}, 359 (2005).}

\bibitem{E852:2004rfa}
    M. Lu \textit{et al.} (E852 Collaboration), \href{https://doi.org/10.1103/PhysRevLett.94.032002}{Phys. Rev. Lett. \textbf{94}, 032002 (2005).}

\bibitem{Baker:2003jh}
    C. A. Baker \textit{et al.}, \href{https://doi.org/10.1016/S0370-2693(03)00643-9}{Phys. Lett. B \textbf{563}, 140--149 (2003).}

\bibitem{CLAS:2008zko}
    M. Nozar \textit{et al.} (CLAS Collaboration), \href{https://doi.org/10.1103/PhysRevLett.102.102002}{Phys. Rev. Lett. \textbf{102}, 102002 (2009).}


\bibitem{ParticleDataGroup:2022pth}
    R. L. Workman \textit{et al.} (Particle Data Group), \href{https://doi.org/10.1093/ptep/ptac097}{Prog. Theor. Exp. Phys. \textbf{2022}, 083C01 (2022).}


\bibitem{Woss:2019hse}
    A. J. Woss, C. E. Thomas, J. J. Dudek, R. G. Edwards, and D. J. Wilson (Hadron Spectrum Collaboration), \href{https://doi.org/10.1103/PhysRevD.100.054506}{Phys. Rev. D \textbf{100}, 054506 (2019).}



\bibitem{OmegaPhoton:1983eln}
    M. Atkinson \textit{et al.} (Omega Photon Collaboration), \href{https://doi.org/10.1016/0550-3213(83)90332-2}{Nucl. Phys. \textbf{B229}, 269 (1983).}

\bibitem{Blackett:1997cm}
    G. R. Blackett, K. Danyo, T. Handler, M. Pisharody, and G. T. Condo, \href{https://arxiv.org/abs/hep-ex/9708032}{arXiv:hep-ex/9708032.}

\bibitem{OmegaPhoton:1986aqd}
    M. Atkinson \textit{et al.} (Omega Photon Collaboration), \href{https://doi.org/10.1007/BF01566756}{Z. Phys. C \textbf{34}, 157 (1987).}

\bibitem{GlueX:2020idb}
    S. Adhikari \textit{et al.} (GlueX Collaboration), \href{https://doi.org/10.1016/j.nima.2020.164807}{Nucl. Instrum. Methods Phys. Res., Sect. A \textbf{987}, 164807 (2021).}


\bibitem{Barbosa:2015bga}
      F. Barbosa, C. Hutton, A. Sitnikov, A. Somov, S. Somov, and I. Tolstukhin \href{https://doi.org/10.1016/j.nima.2015.06.012}{Nucl. Instrum. Methods Phys. Res., Sect. A \textbf{795}, 376 (2015).}

\bibitem{Berdnikov:2015jja}
      V. V. Berdnikov, S. V. Somov, L. Pentchev, and B. Zihlmann \href{https://doi.org/10.1134/S0020441215010030}{Instrum. Exp. Tech. \textbf{58}, 25 (2015).}
      
\bibitem{Dugger:2017zoq}
    M. Dugger \textit{et al.}, \href{https://doi.org/10.1016/j.nima.2017.05.026}{Nucl. Instrum. Methods Phys. Res., Sect. A \textbf{867}, 115 (2017).}

\bibitem{Pentchev:2017omk}
      L. Pentchev, F. Barbosa, V. Berdnikov, D. Butler, S. Furletov, L. Robison, and B. Zihlmann, \href{https://doi.org/10.1016/j.nima.2016.04.076}{Nucl. Instrum. Methods Phys. Res., Sect. A \textbf{845}, 281 (2017).}

\bibitem{Beattie:2018xsk}
    T. D. Beattie \textit{et al.}, \href{https://doi.org/10.1016/j.nima.2018.04.006}{Nucl. Instrum. Methods Phys. Res., Sect. A \textbf{896}, 24 (2018).}

\bibitem{Pooser:2019rhu}
      E. Pooser \textit{et al.}, \href{https://doi.org/10.1016/j.nima.2019.02.029}{Nucl. Instrum. Methods Phys. Res., Sect. A \textbf{927}, 330 (2019).}

\bibitem{Jarvis:2019mgr}
    N. S. Jarvis \textit{et al.}, \href{https://doi.org/10.1016/j.nima.2020.163727}{Nucl. Instrum. Meth. A \textbf{962}, 163727 (2020).}

\bibitem{geant4}
    S. Agostinelli \textit{et al.}, \href{https://doi.org/10.1016/S0168-9002(03)01368-8}{Nucl. Instrum. Methods Phys. Res., Sect. A \textbf{506} 250 (2003).}


\bibitem{Batty:1976cm}
    C. J. Batty, S. D. Hoath, and B. L. Roberts, \href{https://doi.org/10.1016/0029-554X(76)90265-2}{Nucl. Instrum. Methods \textbf{137}, 179 (1976).}

\bibitem{SuppMaterials}
    See Supplemental Material at (link to be created by Journal) for additional information about systematic uncertainties


\bibitem{CONDO:SLAC1993}
    G. T. Condo, T. Handler, W. M. Bugg, G. R. Blackett, M. Pisharody, and K. A. Danyo, \href{https://doi.org/10.1103/PhysRevD.48.3045}{Phys. Rev. D \textbf{48}, 3045 (1993).}

\bibitem{Malte:Hadron}
    M. Albrecht, \href{https://doi.org/10.2172/2280643}{Search for Exotic Hadrons in $\eta^{(\prime)}\pi$ at GlueX, (HADRON 2023 Genova, Italy), 10.2172/2280643.}






\end{thebibliography}


\end{document}


\newcommand{\neutralUL}{135}
\newcommand{\chargedUL}{483}
\newcommand{\neutralULdiv}{194}
\newcommand{\chargedULdiv}{694}
\newcommand{\neutralRatio}{3.17}
\newcommand{\chargedRatio}{1.53}

\title{Supplemental Material}
\date{January 9, 2025}
\maketitle
\tableofcontents

\section{Example Fits to $\pi^+\pi^-\pi^0$ distributions}

\begin{figure*}[htbp!]\centering
\begin{overpic}[scale=0.27]{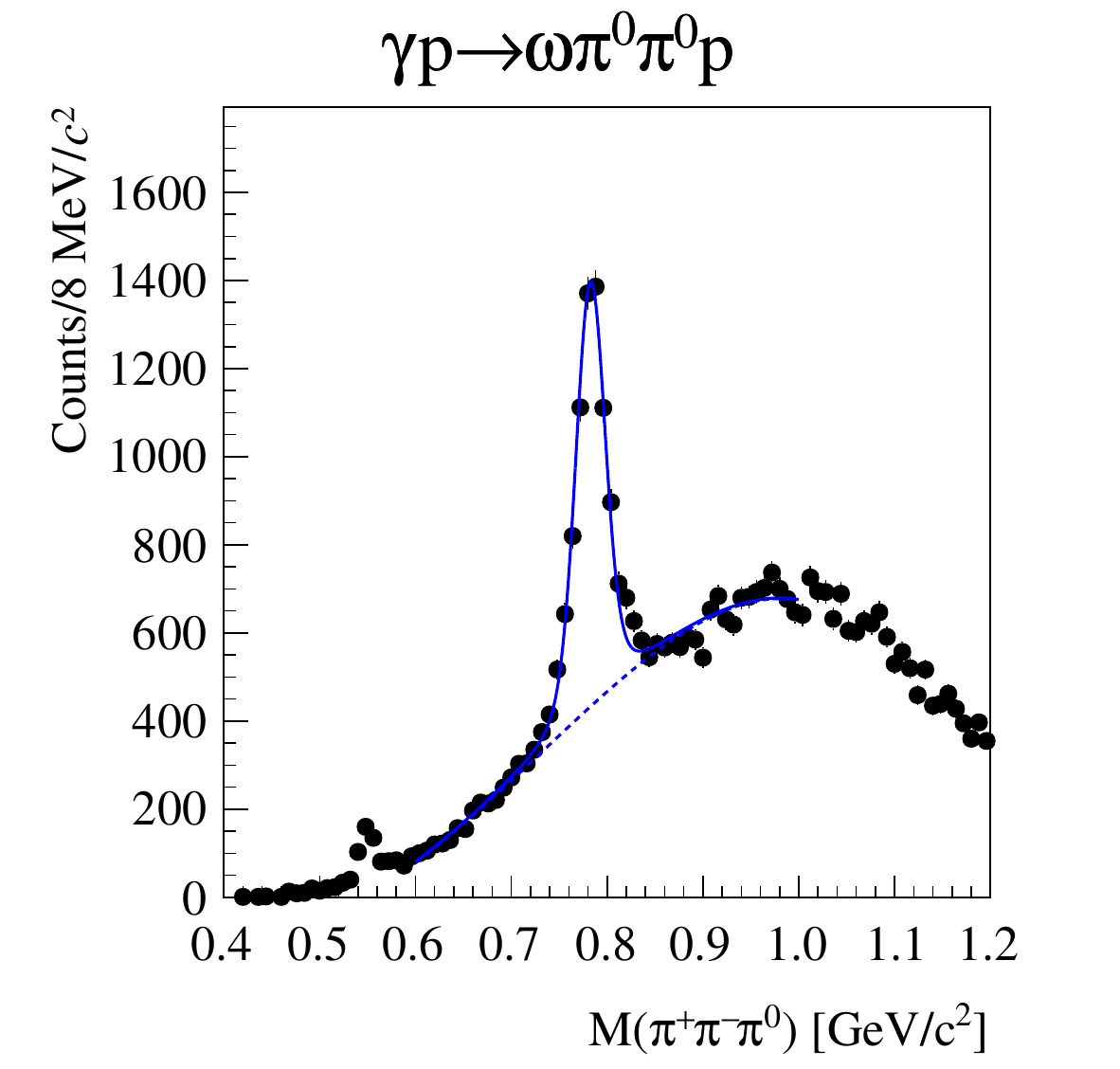}
\put(25,50){(R1)}
\end{overpic}\begin{overpic}[scale=0.27]{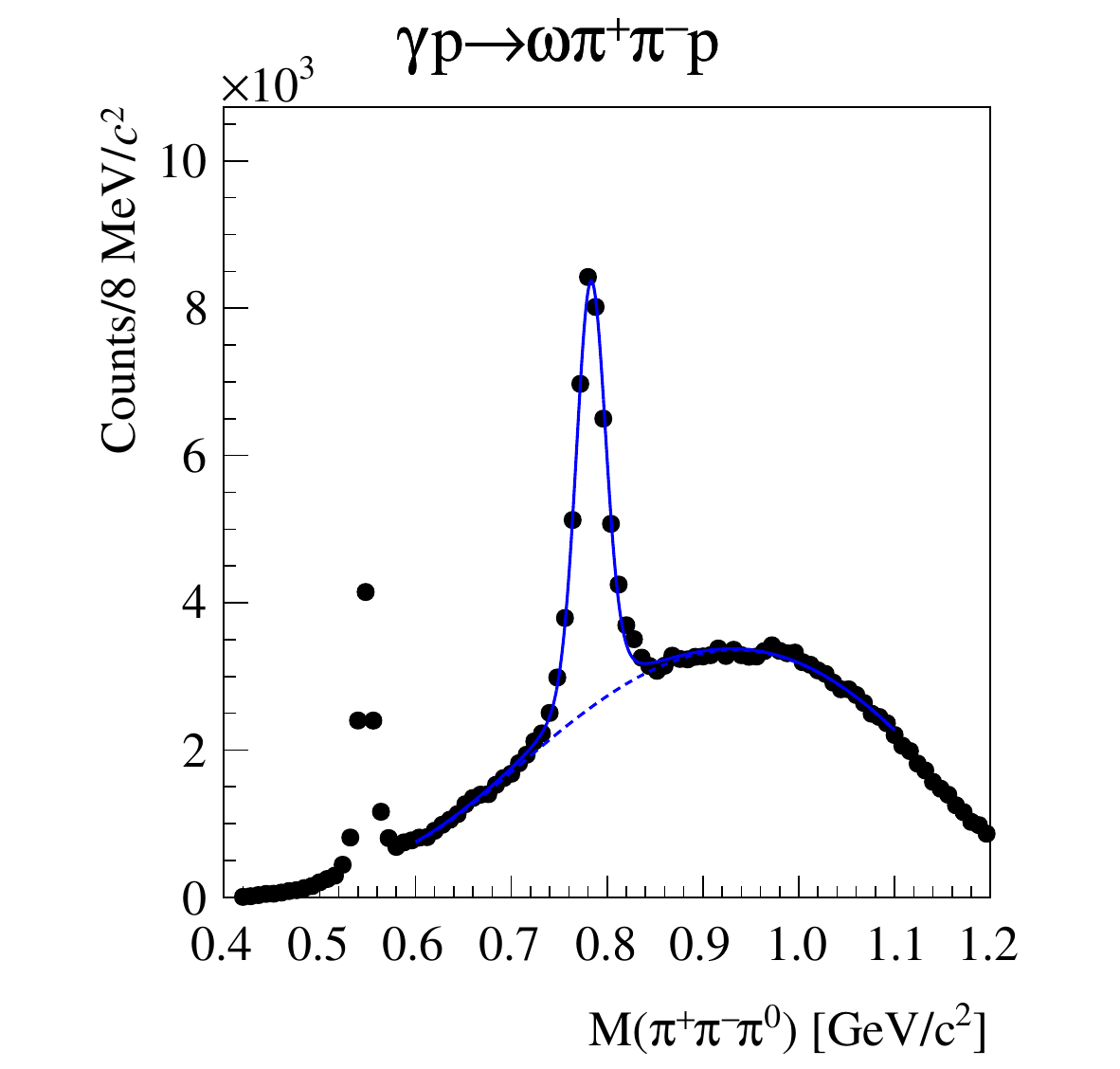}
\put(25,50){(R2)}
\end{overpic}\begin{overpic}[scale=0.27]{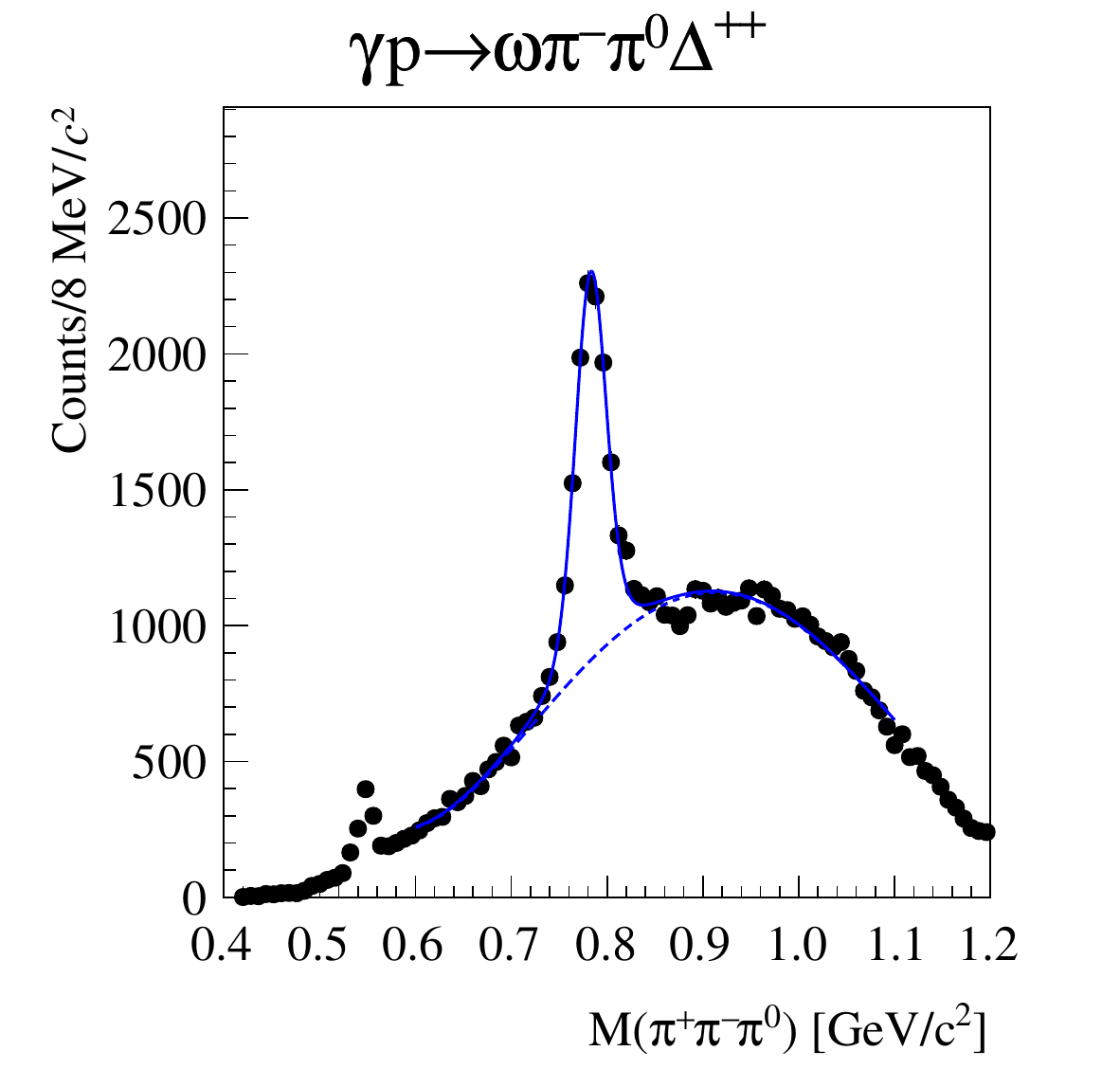}
\put(25,50){(R3)}
\end{overpic}
\caption[]{\label{fig:mass-3pi}The $\pi^{+}\pi^{-}\pi^{0}$ invariant mass distributions in data (black points) fit with a 4th-order polynomial (dashed line) plus a Voigtian (solid line). A clear $\omega(782)$ signal is seen in all three reactions, which is accompanied by a smaller signal for the $\eta(548)$. }
\end{figure*}

Figure \ref{fig:mass-3pi} shows example fits to the $\omega$ candidate masses in data for all three reactions. The data are fit with a fourth-order polynomial to describe the background, plus a Voigtian to describe the $\omega$ signal.  

\section{Measurement of $\sigma(\gamma p\to a_2^-(1320)\Delta^{++})$}

\begin{figure}
    \centering
    \includegraphics[scale=0.3]{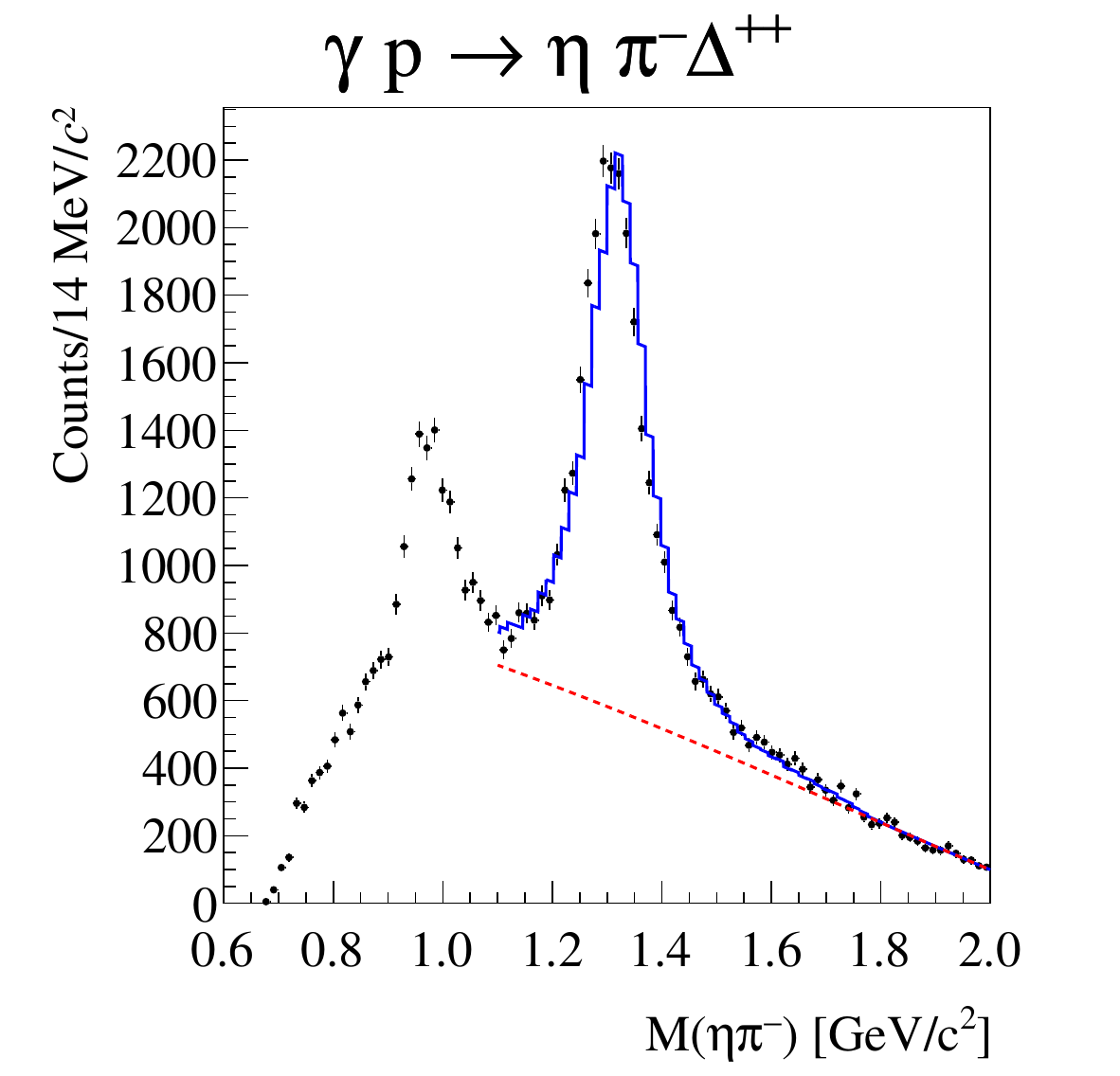}
    \caption{Fit to the invariant mass of $\eta\pi^-$  in $\gamma p\to \eta\pi^-\Delta^{++}$ with $\eta\to\pi^+\pi^-\pi^0$, which is used to estimate the size of $\sigma(\gamma p\to a_2^-(1320)\Delta^{++})$. }
    \label{fig:etapimFit}
\end{figure}

As mentioned in the paper, we use the process $\gamma p\to \eta\pi^-\Delta^{++}$ with $\eta\to\pi^+\pi^-\pi^0$ to estimate the size of $\sigma(\gamma p\to a_2^-(1320)\Delta^{++})$. The selection criteria match those used for $\gamma p\to\omega\pi^-\pi^0\Delta^{++}$. We select a 50~MeV/$c^2$ window around the PDG mass of the $\eta$ in the $\pi^+\pi^-\pi^0$ invariant mass distribution and fit the corresponding $\eta\pi^-$ invariant mass distribution. The goal of these fits is to find the minimum $a_2^-(1320)$ signal consistent with the data, since that will result in the largest upper limit on the ratio $\sigma(\gamma p\to \pi_1^-(1600)\Delta^{++})/\sigma(\gamma p \to a_2^-(1320)\Delta^{++})$. We perform six fit variations to determine the smallest $a_2^-(1320)$ cross section. All fit variations use a signal Monte Carlo shape to describe the $a_2^-(1320)$ signal. We use a third-, fourth-, or fifth-order polynomial to describe the background. We also use two fit ranges: the first uses the mass range from 1.1~GeV$/c^2$ to 2.0~GeV/$c^2$, while the second adds in the region from 0.69~GeV/$c^2$ to 0.90~GeV/$c^2$. The lowest cross section out of these six variations is used to determine the upper limit on the ratio of cross sections. An example fit for the fit range from 1.1~GeV/$c^2$ to 2.0~GeV/$c^2$ with a fourth-order polynomial is shown in Figure~\ref{fig:etapimFit}. This example fit gives $\sigma(\gamma p\to a_2^-(1320)\Delta^{++})=350$~nb, where we do not display any uncertainties because there are large systematic uncertainties, some of which cancel in the ratio of cross sections.

\section{Upper Limit Projections}

In the main body of the paper, we present the largest possible $\pi_1(1600)$ contributions for the $\eta^{(\prime)}\pi^-$ decay modes based on our 90\% C.L. upper limits. In this section, we show details on how the maximum values for $R=\mathcal{B}(\pi_1(1600)\to\eta^{(\prime)}\pi)/\mathcal{B}(\pi_1(1600)\to b_1\pi)$ were determined, describe the selection criteria used for the four reactions, and show the corresponding projection plots for the $\eta^{(\prime)}\pi^0$ decay modes.

To estimate the maximum $\pi_1(1600)$ yield in the $\eta^{(\prime)}\pi$ decay modes, we maximize the value 
\begin{equation}
\begin{split}
    R=\frac{\mathcal{B}(\pi_1(1600)\to \eta^{(\prime)}\pi)}{\mathcal{B}(\pi_1(1600)\to b_1\pi)}\\
    =\frac{\Gamma(\pi_1(1600)\to\eta^{(\prime)}\pi)}{\Gamma(\pi_1(1600)\to b_1\pi)}
    \end{split}
\end{equation}
by making $\Gamma(\pi_1(1600)\to\eta^{(\prime)}\pi)$ as large as possible and $\Gamma(\pi_1(1600)\to b_1\pi)$ as small as possible within the range of values allowed by Table III of Ref.~\cite{Woss:2020ayi}, which corresponds to the full range of values from all of their LQCD fit results. To be consistent with our $\omega\pi\pi$ analysis, we use the same value for the $\pi_1(1600)$ total width used to set the upper limit. This width assumption combined with the range of decay widths given in Table III of Ref.~\cite{Woss:2020ayi} gives the values in Table~\ref{tab:pi1BF_jpac}, where $\Gamma_{\text{max}}(\pi_1(1600)\to\text{others})$ is the maximum partial width for all decay modes other than the $\eta^{(\prime)}\pi$ and $b_1\pi$ decay mode being considered.

\begin{table*}
    \centering
        \caption{The decay widths from Table III of Ref.~\cite{Woss:2020ayi} that maximize the ratio of branching fractions $R$  for $\Gamma_{\text{tot}}(\pi_1(1600))=492$~MeV, the $\pi_1(1600)$ width measured by JPAC~\cite{JPAC:2018zyd}.}
    \begin{tabular}{ccccc}\hline 
       Decay & $\Gamma_{\text{max}}(\pi_1(1600)\to\eta^{(\prime)}\pi)$ & $\Gamma_{\text{max}}(\pi_1(1600)\to\text{others})$ & $\Gamma_{\text{min}}(\pi_1(1600)\to b_1\pi)$ &  $R$    \\\hline
        $\eta\pi$ & 1~MeV & 60~MeV & 431~MeV & 2.3$\times 10^{-3}$\\
        $\eta^{\prime}\pi$ & 12~MeV & 49~MeV &  431~MeV &  2.8$\times 10^{-2}$ \\\hline
    \end{tabular}
    \label{tab:pi1BF_jpac}
\end{table*}

As mentioned in the paper, we select exclusive events for the processes $\gamma p\to \eta^{(\prime)}\pi^0p$ and $\gamma p\to \eta^{(\prime)}\pi^- \Delta^{++}$ with $\Delta^{++}\to\pi^+p$, $\eta'\to\pi^+\pi^-\eta$, and $\eta\to\gamma\gamma$. We only select beam photons in the range $8.2$~GeV$<E_\gamma<8.8$~GeV, which is the beam-energy range that has the maximum photon polarization. The partial-wave analyses of the $\eta^{(\prime)}\pi$ systems will include the beam polarization in the fits to determine the production mechanisms involved. The $\omega\pi\pi$ cross sections do not need the photon polarization information, which allows us to use a wider range of beam energies for those measurements. The $a_2(1320)$ and $\omega\pi\pi$ cross sections are nearly constant over the beam energy range between 8~GeV and 10~GeV. This means there is a negligible systematic uncertainty in projecting the $\pi_1(1600)$ limit determined in the 8~GeV to 10~GeV range to the more narrow 8.2~GeV to 8.8~GeV range. 

For $\gamma p\to \eta^{(\prime)}\pi^-\Delta^{++}$, we use the same selection criteria as for $\gamma p \to \omega\pi^-\pi^0\Delta^{++}$. For $\gamma p\to \eta^{(\prime)}\pi^0p$, we use the same selection criteria as for $\gamma p\to \omega\pi^+\pi^-p$. The $\gamma p\to\eta'\pi^0p$ reaction has two additional backgrounds that require removal. The first comes from the reaction $\gamma p\to \eta'\Delta^{+}$ with $\Delta^{+}\to\pi^0p$. We remove this background process with the requirement $M(\pi^0p)>1.35$~GeV/$c^2$. The second background comes from mismatched photons. Both the $\eta$ and $\pi^0$ decay to two photons. For $\gamma p\to \eta'\pi^0p$, the invariant mass distributions of one photon from the $\eta$ and combined with one photon from the $\pi^0$ show peaks at the $\pi^0$ mass. This background due to miscombined photons is removed by requiring that the mass of one photon from the $\eta$ and one photon from the $\pi^0$ are not within a 30~MeV$/c^2$ wide window of the nominal $\pi^0$ mass. We select the $\eta'$ by requiring $|M_{\pi^+\pi^-\eta}-m_{\eta',\text{PDG}}|<$~25~MeV/$c^2$. We subtract non-$\eta'$ background contributions using $\eta'$ sidebands with $|M_{\pi^+\pi^-\eta}-m_{\eta',\text{PDG}}\pm 60~\textrm{MeV}/c^2|<$~25~MeV/$c^2$ for $\gamma p\to\eta'\pi^-\Delta^{++}$. For $\gamma p\to\eta'\pi^0p$, there is a sharp increase in background events in the $\eta'$ candidate masses at around 1.05~GeV/$c^2$. To avoid this region in our sideband subtraction, we use sidebands defined by $25~\text{MeV/}c^2<|M_{\pi^+\pi^-\eta}-m_{\eta',\text{PDG}}|<$~50~MeV/$c^2$.

\begin{figure*}
    \centering
    \includegraphics[scale=0.55]{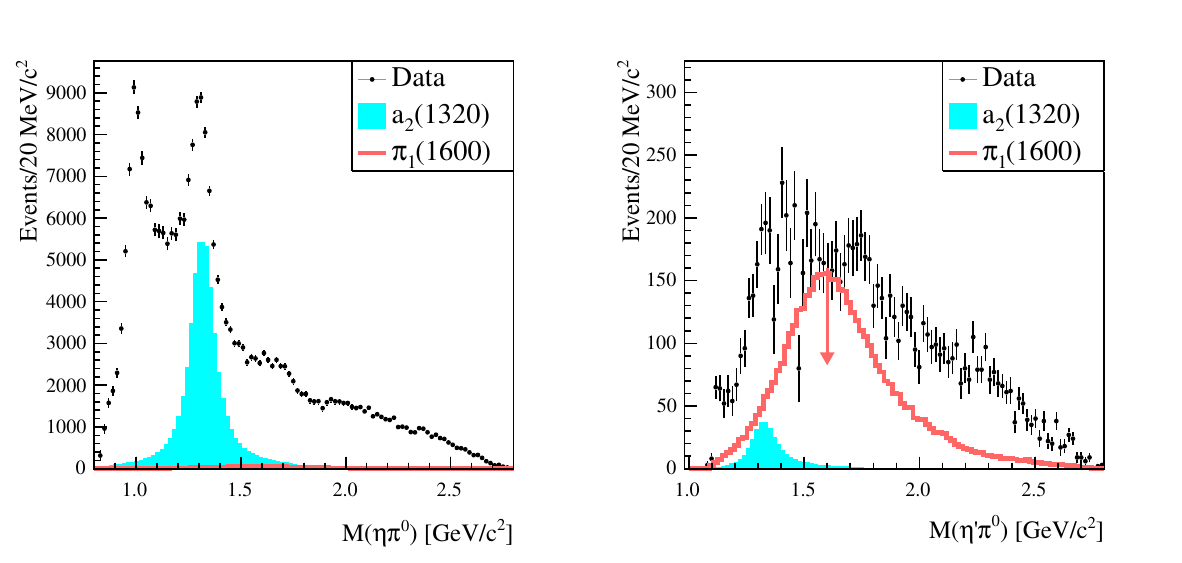}
        \caption{The reconstructed $\eta^{(\prime)}\pi^0$ invariant mass distributions (points), overlaid with the $a_2(1320)$ signal (cyan) and the $\pi_1(1600)$ upper limit (red). }
    \label{fig:neutral}
\end{figure*}

We show the projected $\pi_1(1600)$ upper limits for the $\eta^{(\prime)}\pi^0$ mass distributions in Figure \ref{fig:neutral}. The trends are very similar to the $\eta^{(\prime)}\pi^-$ distributions shown in the main paper: the $\pi_1(1600)$ is expected to be less than a 1\% contribution to the $\eta\pi^0$ system, while it could be the main contribution to $\eta'\pi^0$.

\section{Systematic Uncertainties on $\sigma(\omega\pi\pi)$}

This section details all the systematic uncertainties included in our measurements of the $\omega\pi\pi$ cross sections, with the results summarized in Table~\ref{tab:xsSys}. 

\begin{table}
    \caption{Total systematic uncertainty for each $\omega\pi\pi$ cross section. Not listed is the variation of the $\omega$ fit model, which varies bin-to-bin. }
    \centering
    \begin{tabular}{lccc}\hline
        Source & $\sigma(\omega\pi^+\pi^-)$ & $\sigma(\omega\pi^0\pi^0)$ & $\sigma(\omega\pi^-\pi^0)$ \\\hline
        Track efficiency & 13.5\% & 9.1\% & 16.0\%  \\
       Photon efficiency & 8.1\% & 24.3\% &  16.3\% \\ 
        Luminosity & 5.0\% & 5.0\% & 5.0\% \\
        Kinematic fit & 6.7\% & 6.8\% & 7.1\% \\ 
      Unused shower energy & - & 0.7\% & - \\
        MC substructure & 2.9\% & - & 0.9\%\\ 
        MC angles & 1.5\% & - & 0.9\%\\
        MC $-t$ slope & 2.6\% & 2.4\% & 0.3\%\\
        Proton momentum & 1.7\% & 0.6\% & 1.7\%\\
        Total & 18.4\% & 27.4\% & 24.5\%\\\hline
   
    \end{tabular}
    \label{tab:xsSys}
\end{table}

There are several systematic effects that could potentially bias the overall normalization of our cross section estimates. Systematic uncertainties arise due to a potential mismatch of the charged-track and photon reconstruction efficiencies in data and Monte Carlo (MC). We assign conservative systematic uncertainties of 5\% for the recoil proton, and 3\% (5\%) for each photon that enters the forward calorimeter (barrel calorimeter). For the charged pions, we determine the systematic uncertainty using the method discussed in Section~15.1 of Ref.~\cite{GlueX:2020idb}. Using this method, we determine the ratio of efficiencies estimated from Monte Carlo and real data in bins of polar angle $\theta_\pi$ with respect to the beam axis. The systematic uncertainty is given by how far this quantity deviates from unity. The study has high precision for the region $\theta_\pi < 12^\circ$, so we determine a weighted average systematic uncertainty for pions in this angular region. The study also covers larger polar angles, but with lower precision. We assign conservative systematic uncertainties of 3\% for $12^\circ<\theta_\pi<20^\circ$ and 5\% for $\theta_\pi>20^\circ$. To determine the total tracking or photon efficiency systematic uncertainty for each final state, we assume that the uncertainties for each track and photon are 100\% correlated. When combining 100\% correlated uncertainties, the uncertainties add linearly instead of in quadrature. 

We assign a conservative systematic uncertainty of 5\% for the luminosity measurement, which dominantly comes from our understanding of the pair spectrometer. 

To determine the systematic uncertainty from the kinematic fit, we use data on $\gamma p\to \eta'p$ with $\eta'\to\pi^+\pi^-\eta$ and $\eta'\to\pi^0\pi^0\eta$ with $\eta\to\pi^+\pi^-\pi^0$ as control samples for the $\omega\pi^+\pi^-$ and $\omega\pi^0\pi^0$ data, respectively. This is done because the $\eta'$ provides a narrow peak with well-controlled backgrounds, and has kinematics similar to the $M(5\pi)$ region that we are interested in for our upper limits. To determine the systematic uncertainty from the kinematic fit, we look at how much the $\eta'$ cross section changes as we vary our selection on the $\chi^2/\text{ndf}$ of the kinematic fit in the range $5<\chi^2/\text{ndf}<25$. The difference between the minimum and maximum values of the cross section is used as an estimate for the systematic uncertainty. The systematic uncertainty determined using this method is 6.7\% for $\omega\pi^+\pi^-$, and 6.8\% for $\omega\pi^0\pi^0$. For the $\omega\pi^-\pi^0$ system, there is no such clean control sample. To determine the systematic uncertainty in this case, we directly measure $\sigma(\gamma p\to \omega\pi^-\pi^0\Delta^{++})$ for $1.2$~GeV/$c^2<\nolinebreak m_{\omega\pi\pi}<2.2$~GeV/$c^2$ and vary our selection on the $\chi^2/\text{ndf}$ in the range $5<\chi^2/\text{ndf}<20$. From the difference between the minimum and maximum values, we get a systematic uncertainty of 7.1\%. Note that these are conservative estimates of the systematic uncertainty since at larger $\chi^2/\text{ndf}$, non-exclusive events can inflate the measured cross section. 

For the $\omega\pi^0\pi^0$ analysis, we require that the events have no additional reconstructed calorimeter showers to suppress the background process $\gamma p \to \omega\pi^0\pi^0\pi^0p$, which has an additional $\pi^0$. To estimate the systematic uncertainty from this requirement, we measure the cross section of $\gamma p\to \eta' p$ with $\eta'\to\pi^0\pi^0\eta$ and $\eta\to\pi^+\pi^-\pi^0$ before and after applying this requirement. We find that the cross section changes by 0.7\%, which we assign as a systematic uncertainty.

The nominal signal Monte Carlo simulates $\pi_1(1600)\to\omega\pi\pi$ using a uniform distribution in phase space. To determine the systematic uncertainty due to the decay model, we produce independent MC samples that include a Breit-Wigner for $b_1(1235)\to\omega\pi$ to replicate the substructure of $\pi_1^0(1600)\to b_1^\pm \pi^\mp$ and $\pi_1^-(1600)\to b_1^-\pi^0$ or $\pi_1^-(1600)\to b_1^0\pi^-$. The largest difference in the reconstruction efficiency with respect to the nominal Monte Carlo is taken as the systematic uncertainty, which gives 2.9\% for $\omega\pi^+\pi^-$ and 0.9\% for $\omega\pi^-\pi^0$. 

The nominal MC simulation assumes pure $S$-wave decays for both $\pi_1(1600)\to b_1\pi$ and $b_1\to\omega\pi$. The $\pi_1(1600)\to b_1(1235)\pi$ decay has limited phase space, so we expect the $D$-wave contribution to be kinematically suppressed. To account for the possible presence of $D$-waves in $b_1\to\omega\pi$, we determine the efficiency assuming purely $D$-wave angular distributions for this decay. We include this efficiency by reweighting the MC data using the measured $D/S$ amplitude ratio for the $b_1$ of 0.277 from the PDG \cite{ParticleDataGroup:2022pth}. This gives a systematic uncertainty of 1.5\% for $\omega\pi^+\pi^-$ and 0.9\% for $\omega\pi^-\pi^0$.

The $-t$ distributions in data and Monte Carlo may not precisely match, which could cause a mismatch of the efficiency in data and Monte Carlo. In signal Monte Carlo, all the signal reactions are simulated with a $-t$ slope of $5~($GeV$/c)^{-2}$. As a variation, we use the efficiency-corrected $-t$ distribution obtained from data to reweight our Monte Carlo data. This gives systematic uncertainties of 2.6\% for $\omega\pi^+\pi^-$, 2.4\% for $\omega\pi^0\pi^0$, and 0.3\% for $\omega\pi^-\pi^0$.

Previous studies have shown the efficiencies for low-momentum protons can differ in Monte Carlo and data. To study this effect, we directly measure how much the $\omega\pi\pi$ cross section for the range $1.2$~GeV/$c^2<M(\omega\pi\pi)<2.2$~GeV/$c^2$ changes when we require a minimum proton momentum that is 50 MeV/$c$ larger than in our nominal selection. We get systematic uncertainties of 1.7\%, 0.6\%, and 1.7\% for $\omega\pi^+\pi^-$, $\omega\pi^0\pi^0$, and $\omega\pi^-\pi^0$, respectively.

Another source of uncertainty is the fit model used to extract the $\omega$ yield. To account for this, we vary the $\omega$ signal shape from a Voigtian to shapes that are determined from Monte Carlo data and either include all possible $M(\pi^+\pi^-\pi^0)$ combinations or include only the $\pi^+\pi^-\pi^0$ combination coming from the generated $\omega$ decay. We also vary the background polynomial from order 4 to order 3 or 5. These variations are done independently, so in total there are 9 fit variations. The systematic uncertainty for the measured $\omega\pi\pi$ cross sections are given by the standard deviation of the 9 fit variations. 

The total systematic uncertainties are determined by adding the uncertainties from each source in quadrature. This gives a total systematic uncertainty of 18.4\% for $\sigma(\gamma p \to \omega\pi^+\pi^-p)$, 27.4\% for $\sigma(\gamma p \to \omega\pi^0\pi^0p)$, and 24.5\% for $\sigma(\gamma p \to \omega\pi^-\pi^0 \Delta^{++})$. These systematic uncertainties are included in the error bars of Figure~1 of the main paper.

\section{Systematic Uncertainties on $\sigma(\omega\pi\pi)_{I=1}$}

Before we can determine the upper limit on the $\pi_1(1600)$ cross sections, we need to determine the systematic uncertainties on the isospin-1 $\omega\pi\pi$ cross sections. For the charge-exchange process, this is 24.5\%, as given above. To determine the neutral isospin-1 $\omega\pi\pi$ cross section, we have to take the difference of two cross sections:
\begin{equation}
    \sigma(\omega\pi\pi)_{I=1}=\sigma(\omega\pi^+\pi^-)- 2\sigma(\omega\pi^0\pi^0)
\end{equation}
When determining the isospin-1 cross section, we account for the 3\% background contamination in the $\gamma p\to\omega\pi^0\pi^0p$ reaction by multiplying that cross section by 0.97. The uncertainty on the difference $Z=A-2B=\sigma(\omega\pi\pi)_{I=1}$ of the two cross sections $A=\sigma(\omega\pi^+\pi^-)$ and $B=\sigma(\omega\pi^0\pi^0)$ is given by 
\begin{equation}
    \delta_Z^2=\left(\frac{\partial Z}{\partial A}\right)^2\delta_A^2+\left(\frac{\partial Z}{\partial B}\right)^2\delta_B^2+2\rho_{AB}\frac{\partial Z}{\partial A}\frac{\partial Z}{\partial B}\delta_{A}\delta_{B}.
\end{equation}
with $\delta$ representing absolute uncertainties and $\rho_{AB}$ the Pearson correlation coefficient. As mentioned above, the charged tracking efficiency, photon efficiency, luminosity, and kinematic fit uncertainties are correlated across the three measurements. For these systematics, we use $\rho_{AB}=1$. For all other systematic uncertainties, we assume the sources are uncorrelated ($\rho_{AB}=0$). For the case of correlated uncertainties, we find 
\begin{equation}
    \delta_Z=|2\delta_B-\delta_A|
    \label{eq:corrUnc}
\end{equation}
while for uncorrelated uncertainties, we have 
\begin{equation}
    \delta_Z=\sqrt{\delta_A^2+4\delta_B^2}
    \label{eq:unCorrUnc}
\end{equation}
After determining the systematic uncertainty from each source using Eq.~\ref{eq:corrUnc} and Eq.~\ref{eq:unCorrUnc}, we get the total systematic uncertainty by adding the uncertainty from each source in quadrature. Note that the uncertainty for the neutral isospin-1 $\omega\pi\pi$ cross section depends on the absolute size of the uncertainties, not the relative size. This means we cannot assign a global systematic uncertainty for the neutral $\sigma(\omega\pi\pi)_{I=1}$, but instead must calculate the systematic uncertainty for each mass bin.

\section{Systematic Uncertainties on $\sigma(\pi_1(1600))$}

Finally, to determine the upper limit on the $\pi_1(1600)$ cross section, we use the systematic uncertainties on the isospin-1 cross sections along with additional systematic uncertainties for our modeling of the $a_2(1320)$ and $\pi_1(1600)$.

Three of our largest systematic uncertainties are the photon reconstruction efficiency, charged-track reconstruction efficiency, and luminosity calculation systematic uncertainties. These systematic uncertainties are correlated for each source across the three $\omega\pi\pi$ cross section measurements. To account for these large correlated uncertainties in our $\pi_1(1600)$ upper limits, we rescale our cross sections according to the size of these uncertainties. This is done both for the neutral and charged $\sigma(\omega\pi\pi)_{I=1}$. Our nominal fits shown in the paper use the cross sections before this scaling procedure, but the upper limit is determined after scaling the cross sections such that the isospin-1 cross sections give the largest $\pi_1(1600)$ upper limits. After accounting for the total luminosity, charged-track reconstruction efficiency, and photon reconstruction efficiency uncertainties by rescaling the cross section values, the remaining systematic uncertainties on $\sigma(\gamma p\to \pi_1^-(1600)\Delta^{++})$ listed in Table~\ref{tab:xsSys} still need to be taken into account. This remaining systematic uncertainty is 7.4\%, which comes from adding the remaining uncertainty sources in quadrature.

As mentioned in the paper, we need the probability distribution function of $\sigma_x=\sigma(\pi_1)$ to determine the upper limit. We derive the probability distribution function from the likelihood function of the least-$\chi^2$ fit, which is given by $e^{[\chi^2_{\text{min}}-\chi^2(\sigma_x)]/2}$ , where $\chi^2_\text{min}$ is the $\chi^2$ value of the best fit. To account for the remaining 7.4\% systematic uncertainty ($\delta_{\text{rel}}$) on $\sigma(\gamma p\to \pi_1^-(1600)\Delta^{++})$, we need to convolve this probability distribution function by a Gaussian distribution. We perform this convolution numerically by subdividing the likelihood distribution into 1000 bins of $\sigma(\pi_1)$. For each bin $i$, we replace the uniform distribution of the bin with the Gaussian distribution $N(\sigma_i(\pi_1),\sigma_i(\pi_1)^2\delta_{\text{rel}}^2)$. The sum of all these contributions gives the final probability distribution of $\sigma(\pi_1)$ that incorporates the systematic uncertainties.

The neutral isospin-1 $\omega\pi\pi$ cross section does not have a well-defined total systematic uncertainty, so we cannot account for the systematic uncertainties by convolving the probability distribution with a Gaussian in this case. Instead, we determine the total systematic uncertainty for each $M(\omega\pi\pi)$ bin, and shift the measured value of the cross section up by this total systematic uncertainty when determining the $\pi_1(1600)$ upper limits. The effect is that the probability distribution for the $\pi_1^0(1600)$ cross section is shifted instead of broadened by the systematic uncertainties.

To account for the systematic uncertainty due to the different $\omega$ fit variations listed above, we use the fit variation that gives the largest $\pi_1(1600)$ upper limit. 

In addition, the parameterization of the $a_2(1320)$ and $\pi_1(1600)$ are a potentially large source of uncertainty in the analysis. For the nominal shapes, we model the $a_2(1320)$ using a Breit-Wigner based on the PDG parameters~\cite{ParticleDataGroup:2022pth}, and the $\pi_1(1600)$ using a Breit-Wigner based on the JPAC parameters~\cite{JPAC:2018zyd}. The size of the $a_2(1320)$ contributions are fixed based on the cross sections measured in the $\eta\pi$ systems and the value $\mathcal{B}(a_2(1320)\to\omega\pi\pi)=(10.6\pm 3.2)$\% from the PDG \cite{ParticleDataGroup:2022pth}. We account for the uncertainty in the measured $a_2(1320)$ cross sections by repeating the fits after varying the size of the cross sections up and down by their total uncertainties, obtained by adding the statistical and systematic uncertainties in quadrature. The fits shown in Fig.~2 of the main paper use the nominal $a_2(1320)$ cross sections and branching fractions, while the $\pi_1(1600)$ upper limits are determined from fits where the $a_2(1320)$ cross sections are lowered by their total uncertainty since that gives the largest upper limits. Additionally, we vary the shape of the $\pi_1(1600)$ by changing the mass and width parameters within their uncertainties from Ref.~\cite{JPAC:2018zyd}. We fit with the nominal mass and width, as well as with variations of $\pm 1\sigma_{\text{tot}}$, where $\sigma_{\text{tot}}$ is the total uncertainty from the JPAC measurement. This is done independently for the mass and width, which results in 9 total fit variations of the $\pi_1(1600)$ shape. We report the upper limit from the fit variation that results in the largest upper limit. 

The final source of systematic uncertainty comes from the possible interference between the decays $\pi_1^0(1600)\to b_1^+\pi^-$ and $\pi_1^0(1600)\to b_1^-\pi^+$ or $\pi_1^-(1600)\to b_1^0\pi^-$ and $\pi_1^-(1600)\to b_1^-\pi^0$. To determine the size of the interference effect, we simulate these decay modes assuming totally coherent or incoherent amplitudes. Based on the isospin Clebsch-Gordan coefficients, we expect the two decay modes to destructively interfere. By determining the ratio of MC events for totally destructive interference over the incoherent case, we find that lower $M(\omega\pi\pi)$ values will have more interference, since the $b_1$ bands have more overlap. This means that the size of the interference will vary based on the $\pi_1(1600)$ mass and width that were used in the fit. We find the signal can be reduced by destructive interference by between $20-30\%$ for the $\pi_1(1600)$ in the mass range $1.2$~GeV/$c^2<m_{\omega\pi\pi}<2.2$~GeV/$c^2$. To recover the "true" cross section before this $\pi_1(1600)\to b_1\pi$ interference, we scale the probability function by 1.0/0.7 for the case of 30\% destructive interference.

\section{Error Propagation on Ratio of Cross Sections}

In the main paper, we report upper limits on the ratio of the $\pi_1(1600)$ to $a_2(1320)$ cross sections. To set these upper limits, we need to incorporate the statistical and systematic uncertainties in both the numerator and denominator.

For the $a_2^0(1320)$, we use the measured statistical and systematic cross section uncertainties from the $\eta\pi^0$ PWA result~\cite{Malte:Hadron}. For the $a_2^-(1320)$, we determine the dominant systematic uncertainties and incorporate those into the upper limit. We perform several fit variations to the $\eta\pi^-$ invariant mass distribution, and use the fit variation that gives the smallest value for the $a_2^-(1320)$ cross section. In addition to the fitting systematic uncertainty, we also determine the tracking efficiency (15.3\%) and photon efficiency (7.5\%) systematic uncertainties, as well as the systematic uncertainty due to the kinematic fit (3.6\%).

The photon and charged-track efficiency systematic uncertainties are correlated between the numerator and denominator, so they will partially cancel in the ratio of cross sections. The tracking systematic uncertainty can be broken into two parts, one due to the charged pions $\delta(\pi)$, and one due to the recoil protons $\delta(p)$. The recoil protons from the $\pi_1(1600)$ can populate a wider range of angles than the recoil protons from the $a_2(1320)$, so to be conservative, we assume the proton tracking uncertainties are uncorrelated in this ratio. The charged-pion tracking systematic uncertainties are fully correlated, since they take into account the angular distribution already. Therefore, the total charged-pion tracking systematic uncertainty for the ratio of cross sections is given by $\delta_{\text{ratio}}(\pi) =|\delta_{\text{num}}(\pi)-\delta_{\text{denom}}(\pi)|$. Since we are assuming the proton uncertainties are uncorrelated, the total proton tracking uncertainty is given by $\delta_{\text{ratio}}(p) = \sqrt{\delta_{\text{num}}(p)^2+\delta_{\text{denom}}(p)^2}$. The total tracking uncertainty for the ratio of cross sections is given by the quadrature sum of $\delta_{\text{ratio}}(\pi)$ and $\delta_{\text{ratio}}(p)$. The photon efficiency systematic uncertainties are fully correlated, so we just use the difference between the numerator and denominator. The luminosity systematic uncertainty fully cancels in the ratio of cross sections. 

The tracking and photon systematic uncertainties are all included in the smearing of the probability distribution for the $\pi_1(1600)$. The $a_2(1320)$ probability distribution includes the kinematic fit systematic uncertainty for the $a_2(1320)$ cross section measurement. 

To determine the probability distribution for the ratio of cross sections, we randomly sample the probability distributions for the $\pi_1(1600)$ and $a_2(1320)$ cross sections a million times, recording the value for the ratio of cross sections each time. Based on the cross section ratio distribution, we can calculate the probability of obtaining a particular value for the cross section ratio. We use this probability distribution to determine the 90\% C.L. upper limit on the cross section ratio.